\documentclass[showpacs,prl,onecolumn,aps,superscriptaddress,preprintnumbers,nofootinbib]{revtex4}
\usepackage[T1]{fontenc}
\usepackage[latin9]{inputenc}
\usepackage{graphicx,hyperref}

\usepackage{amsmath,amssymb}
\usepackage{epsfig}
\usepackage{graphicx}
\usepackage{amsmath}
\usepackage{amsfonts}
\def\slashchar#1{\setbox0=\hbox{$#1$}     		
   \dimen0=\wd0                                 	
   \setbox1=\hbox{/} \dimen1=\wd1               	
   \ifdim\dimen0>\dimen1                        	
      \rlap{\hbox to \dimen0{\hfil/\hfil}}      	
      #1                                        	
   \else                                        	
      \rlap{\hbox to \dimen1{\hfil$#1$\hfil}}   	
      /                                         	
   \fi}

\renewcommand{\vec}{\boldsymbol}
\newcommand{\beq}{\begin{equation}}
\newcommand{\eeq}{\end{equation}}
\newcommand{\bea}{\begin{eqnarray}}
\newcommand{\eea}{\end{eqnarray}}
\newcommand{\baa}{\begin{array}}
\newcommand{\eaa}{\end{array}}

\def\eq#1{{Eq.~(\ref{#1})}}

\def\fig#1{{Fig.~\ref{#1}}}

\newcommand{\bas}{\bar{\alpha}_S}
\newcommand{\as}{\alpha_S}

\newcommand{\nn}{\nonumber}

\newcommand{\h}{\frac{1}{2}}

\newcommand{\x}{\vec{x}}

\newcommand{\Lb}{\left(}
\newcommand{\Rb}{\right)}

\renewcommand{\vec}[1]{\boldsymbol{#1}}
\newcommand{\dif}{\mathrm{d}}

\begin{document}
\title{ Non-linear equation in the re-summed next-to-leading order of perturbative QCD: \\the leading twist approximation}
\author{Carlos Contreras}
\email{carlos.contreras@usm.cl}
\affiliation{Departamento de F\'isica, Universidad T\'ecnica Federico Santa Mar\'ia,  Avda. Espa\~na 1680, Casilla 110-V, Valpara\'iso, Chile}
\author{ Eugene ~ Levin}
\email{leving@tauex.tau.ac.il, eugeny.levin@usm.cl}
\affiliation{Departamento de F\'isica, Universidad T\'ecnica Federico Santa Mar\'ia,  Avda. Espa\~na 1680, Casilla 110-V, Valpara\'iso, Chile}
\affiliation{Centro Cient\'ifico-
Tecnol\'ogico de Valpara\'iso, Avda. Espa\~na 1680, Casilla 110-V, Valpara\'iso, Chile}
\affiliation{Department of Particle Physics, School of Physics and Astronomy,
Raymond and Beverly Sackler
 Faculty of Exact Science, Tel Aviv University, Tel Aviv, 69978, Israel}
\author{Rodrigo Meneses}
\email{rodrigo.meneses@uv.cl}
\affiliation{Escuela de Ingenier\'\i a Civil, Facultad de Ingenier\'\i a, Universidad de Valpara\'\i so, General Cruz 222, Valpara\'\i so, Chile}
\author{Michael Sanhueza}
\email{michael.sanhueza.roa@gmail.com}
\affiliation{Departamento de F\'isica, Universidad T\'ecnica Federico Santa Mar\'ia,   Avda. Espa\~na 1680, Casilla 110-V, Valpara\'iso, Chile}
\date{\today}

\begin{abstract}
In this paper, we use the re-summation procedure, suggested in
 Refs.\cite{DIMST,SALAM,SALAM1,SALAM2}, to fix the BFKL kernel
 in the NLO. However, we suggest a different way to introduce the
 non-linear corrections in the saturation region,  which is based
 on  the leading twist non-linear equation. In the kinematic
 region: $\tau\,\equiv\,r^2 Q^2_s(Y)\,\leq\,1$ , where $r$  denotes
 the size of the dipole, $Y$  its rapidity and $Q_s$  the
 saturation scale, we found  that the re-summation contributes
 mostly to the leading twist of the BFKL equation.   Assuming
 that 
 the scattering amplitude is small, we suggest  using  the linear 
evolution
 equation in this region.
  For $\tau \,>\,1$ we are dealing with the re-summation of $\Lb \bas
 \,\ln \tau\Rb^n$ and other corrections in NLO approximation for the
 leading twist.
We find the BFKL kernel in this kinematic region and write the non-linear
 equation, which we solve analytically. We believe the new equation could
 be a basis for a consistent phenomenology based on the CGC approach.

  \end{abstract}

\maketitle

\vspace{-0.5cm}
\tableofcontents






\section{ Introduction}

The Colour Glass Condensate(CGC) approach is the only candidate for an
 effective theory at high energies, which is based on our microscopic
 theory: QCD (see Ref.\cite{KOLEB} for a review). However, it has been
 known for a long time, that to describe the scattering amplitude in the
 framework of CGC\cite{JIMWLK1,JIMWLK2,JIMWLK3,JIMWLK4,JIMWLK5,JIMWLK6,BK}
 we need to include at least  the next-to-leading order corrections to the
 non-linear equations.
 Indeed, the two essential parameters, that determine the high energy
 scattering,  are the BFKL Pomeron\cite{BFKL} intercept, which is equal
 to $ \,2.8 \bas$, which leads to the energy behaviour of the scattering
 amplitude $N \propto \exp\left(2.8\,\bas \ln(\frac{1}{x})\right)$, and 
the
 energy behaviour of the new dimensional scale: saturation momentum  $Q^2_s
 \propto \exp\left(4.88\,\bas \ln(\frac{1}{x})\right)$ . Both  show the 
increase in the leading order CGC approach, which cannot be reconciled 
with the available experimental data. So, the large NLO corrections appear
  as the only way out,  now as well as  two decades ago.
 
 The non-linear equations in the NLO has been written in
 Refs.\cite{NLOBK0,NLOBK01,NLOBK1,NLOBK2,JIMWLKNLO1,
JIMWLKNLO2,JIMWLKNLO3,DIMST}, but their use in high energy phenomenology
 is marred by the instabilities  due to the presence of large and negative
NLO corrections enhanced by double collinear logarithms (see Ref.\cite{DIMST}
 for discussions and references).  These problems have been found in 
Refs.\cite{BFKLNLO,BFKLNLO1}, identified  and solved in Refs.\cite{
SALAM,SALAM1,SALAM2} for the linear BFKL equation. It turns out that
 instabilities are closely related to the  wrong choice of the
 energy(rapidity) scale, and we need to introduce the re-summed NLO
 corrections to cure this problem. 
 
 In this paper, we use the re-summation procedure, suggested in
 Refs.\cite{SALAM,SALAM1,SALAM2}, to fix the BFKL kernel in the NLO,
 which coincides with the kernel used in Ref.\cite{DIMST}. However, 
we suggest a different way to introduce the non-linear corrections, 
than in Ref.\cite{DIMST}, which is based on the
 approach, developed in Ref.\cite{LETU} for the leading twist non-linear
 equation in the LO. Our first observation is that the re-summation
 contributes mostly to the leading twist of the BFKL equation in the
 kinematic region $\tau\,\equiv\,r^2 Q^2_s(Y)\,<\,1$ , where $r$ 
denotes 
 the size of the dipole, $Y$  its rapidity and $Q_s$  the saturation
 scale. We assume that at $\tau=1$ the scattering amplitude is small and
 we can neglect the non-linear corrections. Therefore, for $\tau < 1$ we
 can restrict ourselves by the linear BFKL equation. For $\tau \,>\,1$ we
 are dealing with the re-summation of $\Lb \bas \,\ln \tau\Rb^n$ and other
 corrections in NLO approximation for the leading twist.
 In this paper we find the BFKL kernel in this kinematic region, and
 write the non-linear equation. We found the analytical solution of
 this equation, and believe the new equation could be a basis for a
 consistent phenomenology based on the CGC approach.

\section{  BK non -linear equation}
The BK evolution equation  for the dipole-target scattering amplitude
 $N\Lb \vec{x}_{10},\vec{b},Y ; R\Rb$ has  in the leading
 order (LO)  of perturbative QCD\cite{KOLEB,BK,GLR,MUQI,MV} the following 
form:
\bea \label{BK}
&&\frac{\partial}{\partial Y}N\Lb \vec{x}_{10}, \vec{b} ,  Y; R \Rb = \nn\\
&&\bas\!\! \int \frac{d^2 \vec{x}_2}{2\,\pi}\,K\Lb \vec{x}_{02}, \vec{x}_{12}; \vec{x}_{10}\Rb \Bigg(N\Lb \vec{x}_{12},\vec{b} - \h \vec{x}_{20}, Y; R\Rb + 
N\Lb \vec{x}_{20},\vec{b} - \h \vec{x}_{12}, Y; R\Rb - N\Lb \vec{x}_{10},\vec{b},Y;R \Rb\nn\\
&&-\,\, N\Lb \vec{x}_{12},\vec{b} - \h \vec{x}_{20}, Y; R\Rb\,N\Lb \vec{x}_{20},\vec{b} - \h \vec{x}_{12}, Y; R\Rb\Bigg)
\eea
where $\vec{x}_{i k}\,\,=\,\,\vec{x}_i \,-\,\vec{x}_k$  and $ \vec{x}_{10}
 \equiv\,\vec{r}$, $\vec{x}_{20}\,\equiv\,\vec{r}' $ and $\vec{x}_{12}
 \,\equiv\,\vec{r}\,-\,\vec{r}'$.  $Y$ is the rapidity of the scattering
 dipole and $\vec{b}$ is the impact factor. $K\Lb \vec{x}_{02}, \vec{x}_{12};
 \vec{x}_{10}\Rb$ is the kernel of the BFKL equation which in the leading
 order has the following form:
\beq \label{KERLO}
K^{\rm LO} \Lb \vec{x}_{02}, \vec{x}_{12}; \vec{x}_{10}\Rb\,\,= 
\,\,\frac{x^2_{10}}{x^2_{02}\,x^2_{12}}
\eeq
In \eq{BK} $R$ denotes  the size of the target dipole and $\bas
 \,=\,N_c \as/\pi$ where $N_c$ is the number of colours.

 For the kernel of the LO BFKL equation (see \eq{KERLO}) the eigenvalues
  take the form\cite{BFKL,LIP}:

\beq \label{CHI}
\omega_{\rm LO}\Lb \bas, \gamma\Rb\,\,=\,\,\bas\,\chi^{LO}\Lb \gamma
 \Rb\,\,\,=\,\,\,\bas \Lb 2 \psi\Lb 1\Rb \,-\,\psi\Lb \gamma\Rb\,-\,\psi\Lb
 1 - \gamma\Rb\Rb
\eeq
where $\psi(z)$  denotes the Euler psi-function $\psi\Lb z\Rb =
 d \ln \Gamma(z)/d z$.

In the next-to-leading order (NLO), the non-linear equation
 has  a more complicated form\cite{NLOBK0,NLOBK01,NLOBK1,NLOBK2}:

\begin{align}
 \label{NLOBK}
 \hspace*{0cm}
	\frac{\dif S_{10}}{\dif Y}\!= &\frac{\bas}{2\pi}\!
	\int\!\! \dif^2 x_{2}
	\frac{x_{10}^2}{x_{12}^2 x_{02}^2}
	\Bigg\{ 1 \!+\! \bas b
	\left(\ln x_{10}^2 \mu^2 
 \!-\!\frac{x_{12}^2 \!-\! x_{20}^2}{x_{10}^2}
 \ln \frac{x_{12}^2}{x_{02}^2}\right)
 \nn\\
 & \hspace*{-1cm} 
 + \bas \left(\frac{67}{36} \!-\! \frac{\pi^2}{12} \!-\! 
 \frac{5}{18}\,\frac{N_f}{N_c}
 \!-\! \frac{1}{2}\ln \frac{x_{12}^2}{x_{10}^2} 
 \ln \frac{x_{02}^2}{x_{10}^2}
 \right) \Bigg\}
 \left(S_{12} S_{20} \!-\! S_{10} \right)
\nn\\
 & \hspace*{-1cm}+ \frac{\bas^2}{8\pi^2}
 \int \frac{\dif^2 x_2 \,\dif^2 x_3}{x_{23}^4}
 \Bigg\{-2
 + \frac{x_{12}^2 x_{03}^2 + x_{13}^2  x_{02}^2
 - 4 x_{10}^2 x_{23}^2}{x_{12}^2 x_{03}^2 - x_{13}^2 x_{03}^2}
 \ln \frac{x_{12}^2 x_{03}^2}{x_{14}^2 x_{02}^2}
 \nn\\
 & \hspace*{-1.0cm}
 +\frac{x_{10}^2 x_{23}^2}{x_{12}^2 x_{03}^2}
 \left(1 + \frac{x_{10}^2 x_{23}^2}{x_{12}^2 x_{23}^2 - x_{13}^2 x_{02}^2} \right)
 \ln \frac{x_{12}^2 x_{23}^2}{x_{13}^2 x_{02}^2}
 \Bigg\}
 \left(S_{12} S_{23} S_{03}- S_{12} S_{30} \right)
\end{align}
In \eq{NLOBK} $x_{ik} \,=\,\vec{x}_i \,-\,\vec{x}_j$, $\mu$ denotes
 the renormalization scale for the running QCD coupling, $b
 \,=\,\frac{11 N_c \,-\,2 N_f}{12\,\pi}$ and $N_f$ and $N_c$ are
 the number of fermions and colours, respectively.  $S_{ij}$ denotes
 the S-matrix for   scattering of a dipole of size $x_{i j}$  with
 the target, which can be written using the scattering amplitude
 $N_{ij}$ ,  as follows  $S_{ij} = 1 - N_{ij}$. \eq{NLOBK} gives
 the explicit form of the BFKL kernel in the NLO, but as it has
 been alluded  to  we need to re-sum the NLO corrections to avoid
 instabilities.  We  re-sum  in the approximation,
 that was suggested in Ref.\cite{DIMST}, which we will discuss below. 
It turns out, that in the framework of this re-summation we can neglect
 the contribution, which is proportional to $S_{13}\,S_{32}\,-\,
S_{13} S_{32}$ in \eq{NLOBK}; and reduce \eq{NLOBK} to \eq{BK} with the
 kernel, which has to be found in the re-summed NLO. An additional 
argument
 for such simplification stems from the fact that deep in the saturation
 region,  where $S_{ij} \to 0$,  all terms except the first one,  which
 is proportional to $S_{12}$, are small and can be neglected. In other
 words, deep in the saturation region \eq{NLOBK} reduces to \eq{BK}
 without addressing the specific form of re-summation.

In the next-to-leading order the kernel is derived in
 Refs.\cite{BFKLNLO,BFKLNLO1} and has the following form:
\beq \label{KERNLO1}
\omega_{\rm NLO}\Lb \bas,  \gamma\Rb\,\,=\,\,\bas\,\chi^{LO}\Lb \gamma \Rb\,\,+\,\,\bas^2\,\chi^{NLO}\Lb  \gamma\Rb
\eeq
The explicit form of $\chi^{NLO}\Lb  \gamma\Rb$ is given in
 Ref.\cite{BFKLNLO}. However, $\chi^{NLO}\Lb  \gamma\Rb$ turns
 out to be singular at $\gamma \to 1$,  $\chi^{NLO}\Lb\gamma\Rb
 \,\propto\,1/(1 - \gamma)^3$. Such singularities indicate, that
 we have to calculate higher order corrections to obtain a reliable result.
  The procedure to re-sum high order corrections  is suggested in Ref.
 \cite{SALAM,SALAM1,SALAM2,KMRS}. The resulting spectrum of the BFKL
 equation in the NLO,  can be found from the solution of the following
 equation \cite{SALAM,SALAM1,SALAM2}
\beq \label{KERNLOR}
\omega_{\rm NLO}\Lb \bas,  \gamma\Rb\,=\,\bas \Lb \chi_0\Lb\omega_{\rm NLO}, \gamma\Rb\,+\,\omega_{\rm NLO} \,\frac{\chi_1\Lb \omega_{\rm NLO}, \gamma\Rb}{ \chi_0\Lb\omega_{\rm NLO}, \gamma\Rb}\Rb
\eeq
where
\beq \label{CHI0}
\chi_0\Lb\omega, \gamma\Rb\,\,=\,\,\chi^{LO}\Lb \gamma\Rb \,-\,\frac{1}{ 1 \,-\,\gamma}\,+\,\frac{1}{1\,-\,\gamma\,+\,\omega}
\eeq
and
\bea \label{CHI1}
&&\chi_1\Lb\omega, \gamma\Rb\,\,=\\
&&\,\,\chi^{NLO}\Lb \gamma\Rb\,+\,F\Lb \frac{1}{1 - \gamma}\,-\,\frac{1}{1\,-\,\gamma\,+\,\omega}\Rb\,+\,\frac{A_T\Lb \omega\Rb \,-\,A_T\Lb 0 \Rb}{\gamma^2} \,+\, \frac{A_T\Lb \omega\Rb - b}{\Lb 1\,-\,\gamma\,+\,\omega\Rb^2}\,-\,\frac{A_T\Lb 0\Rb - b}{\Lb 1\,-\,\gamma\Rb^2}\nn
 \eea
Functions $\chi^{NLO}\Lb \gamma\Rb$ and $A_T\Lb \omega\Rb$ as well as 
 the constants ($F$ and $b$),  are defined in 
Refs.\cite{SALAM,SALAM1,SALAM2}.

In Ref. \cite{KMRS} Khoze, Martin, Ryskin and Stirling (KMRS) sugested
   an economic  form of $\chi_1\Lb \omega,\gamma\Rb$,  which coincides
  with \eq{CHI1} to within  $7\%$, and, therefore, gives  reasonable
 estimates of all constants and functions in 
\eq{CHI1}.  The equation for $\omega$  takes the form
\beq \label{KMRSOM}
\omega^{\rm KMRS} \,=\,\bas\Lb 1 - \omega^{\rm KMRS}\Rb \Lb \frac{1}{\gamma}  + \frac{1}{1 - \gamma + \omega^{\rm KMRS}}\,+\,\underbrace{\Lb 2 \psi(1) - \psi\Lb 2 - \gamma\Rb -  \psi\Lb 1 + \gamma\Rb\Rb}_{\mbox{ high twist contributions}}\Rb
\eeq
One  can see that $\gamma(\omega) \to 0$ when $\omega \to 1$ as follows
 from energy conservation.

\section{  NLO BFKL kernel in the  perturbative QCD region}
\subsection{Eigenfunctions of the BFKL equation}
Therefore, the linear BFKL equation in the NLO takes the form:

\beq \label{BFKL}
\hspace{-0.2cm}\frac{\partial}{\partial Y}N\Lb \vec{x}_{10},\vec{b} ,  Y; R \Rb = 
\bas\!\! \int \frac{d^2 \vec{x}_2}{2\,\pi}\,K\Lb \vec{x}_{02}, \vec{x}_{12}; \vec{x}_{10}\Rb \Bigg\{N\Lb \vec{x}_{12},\vec{b} - \h \vec{x}_{20}, Y; R\Rb + 
N\Lb \vec{x}_{20},\vec{b} - \h \vec{x}_{12}, Y; R\Rb - N\Lb \vec{x}_{10},\vec{b},Y; R \Rb\Bigg\}
\eeq
The general solution to \eq{BFKL}  can be written as follows
 \beq \label{GENSOLL}
 N\Lb  r ,  b,  Y; R\Rb\,\,\,=\,\,\,\int^{\epsilon + i \infty}_{\epsilon - i \infty}\frac{d \gamma}{2\,\pi\,i} e^{\omega\Lb \bas,\gamma\Rb \,Y}\, \phi_\gamma\Lb \vec{r} , \vec{R}, \vec{b}\Rb \,\phi_{\rm in}\Lb \gamma,R\Rb
 \eeq
where $ \phi_\gamma\Lb \vec{r} , \vec{R}, \vec{b}\Rb$ is the eigenfunction
 of the BFKL equation and 
$\phi_{\rm in}\Lb \gamma, R\Rb$  can be found from the initial condition at
 $Y = 0$. In \eq{GENSOLL}
 $\vec{r} \,\equiv\,\vec{x}_{10}$ denotes the size of the scattering dipole,
 while $R$  the size of the target. 
   
 In Ref.\cite{LIP} it was proved that the eigenfunction of the BFKL
 equation has the following form

\bea 
\phi_\gamma\Lb \vec{r} , \vec{R}, \vec{b}\Rb\,&=&\,\Lb \frac{ r^2\,R^2}{\Lb \vec{b}  + \h(\vec{r} - \vec{R})\Rb^2\,\Lb \vec{b}  -  \h(\vec{r} - \vec{R})\Rb^2}\Rb^\gamma\,\,=\,\,\,e^{\gamma\,\xi} \label{EIGENF}\\
\mbox{with}~~~ \xi\,&=&\,\ln\Lb \frac{ r^2\,R^2}{\Lb \vec{b}  + \h(\vec{r} - \vec{R})\Rb^2\,
\Lb \vec{b}  -  \h(\vec{r} - \vec{R})\Rb^2}\Rb \label{XI}
\eea
for any kernel,  which satisfies the conformal symmetry. Using 
 \eq{EIGENF} we can re-write the general solution in the form:
\beq \label{GENSOL1}
 N\Lb  \xi ,  Y; R\Rb\,\,\,=\,\,\,\int^{\epsilon + i \infty}_{\epsilon - i \infty}\frac{d \gamma}{2\,\pi\,i} e^{\omega\Lb \bas,\gamma\Rb \,Y \,\,+\,\,\gamma\,\xi} \,\phi_{\rm in}\Lb \gamma, R\Rb  
 \eeq
  
 where $\xi \,=\,\ln\Lb r^2\,Q^2_s\Lb Y=0; \vec{b},\vec{R}\Rb\Rb$.
 Comparing \eq{XI} with this definition  of $\xi$, one can see that 
 \beq \label{XI1} 
 Q^2_s\Lb Y=0; \vec{b},\vec{R}\Rb \,=\, \frac{R^2}{\Lb \vec{b} \, +\,
 \h\vec{R}\Rb^2\,
\Lb \vec{b}\,  -\, \h  \vec{R}\Rb^2}
\eeq

 for $r \ll R$.
\subsection{Eigenvalues of the BFKL equation}

As has been mentioned,  the re-summation, that has been suggested in
 Ref.\cite{DIMST}  is  determined by
the anomalous dimension $\gamma$ in 
 the vicinity of the eigenvalues at $\gamma \,\to\,1$.  The singular
 part of the general kernel in the NLO (see \eq{KERNLOR})  has the
 following form:
\beq \label{GANLOS}
\omega\,\,=\,\,\frac{\bas}{1 - \gamma + \omega};
\eeq
with the solution:
\beq \label{OMNLO}
\omega\Lb \gamma\Rb\,\,=\,\,\h\Lb-\Lb 1  - \gamma\Rb\,+\,\sqrt{4\,\bas \,+\,\Lb 1 - \gamma\Rb^2}\Rb
\eeq

It is instructive to note that \eq{KMRSOM} gives
\beq \label{OMNLOKMRS}
\omega\,\,=\,\,\frac{\bas}{1 - \gamma + \omega}\,\Lb 1\,-\,\omega\Rb; 
\eeq

\begin{figure}
\begin{tabular}{c  c  c}
      \includegraphics[width=7cm]{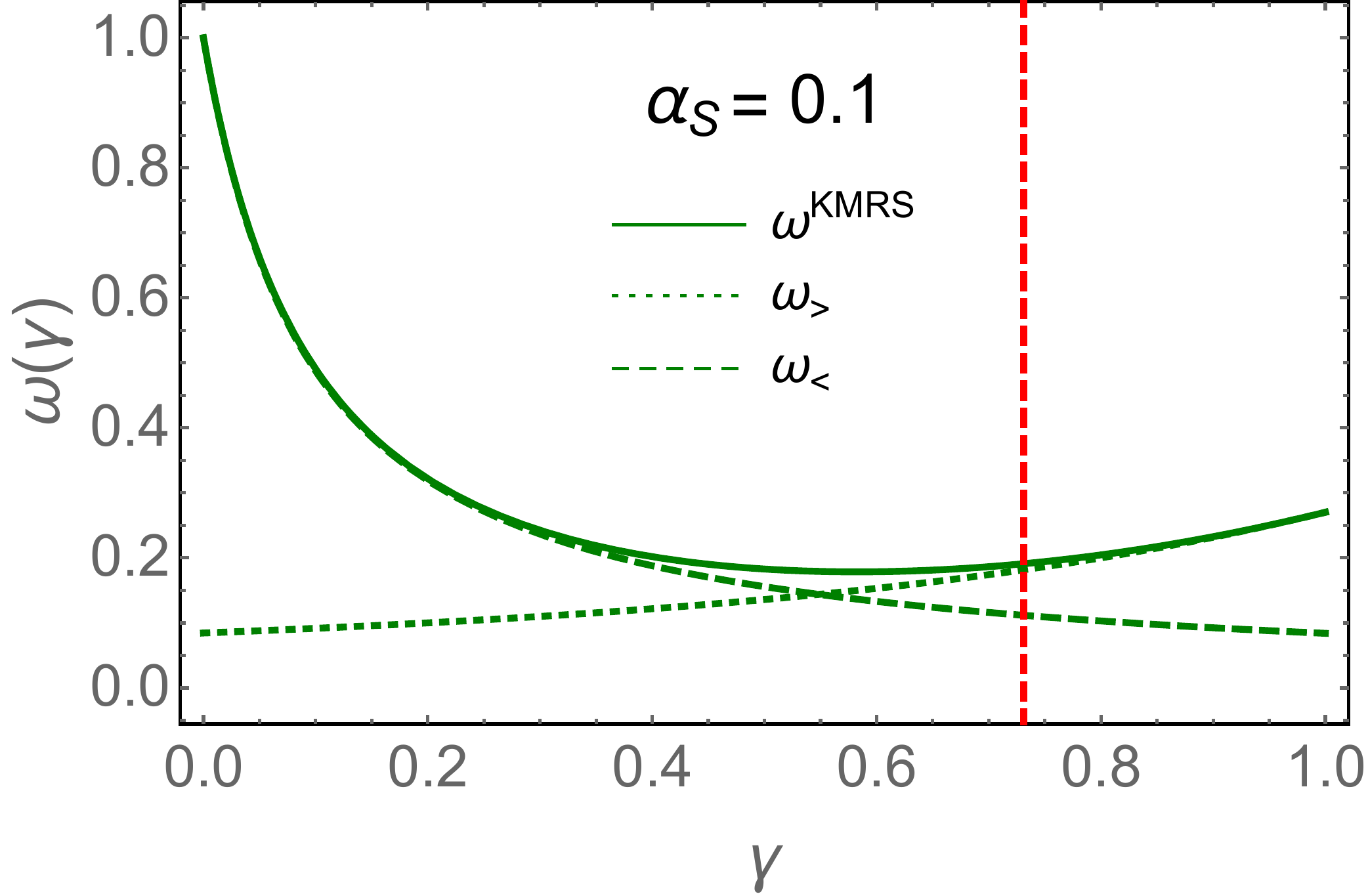}&  & \includegraphics[width=7cm]{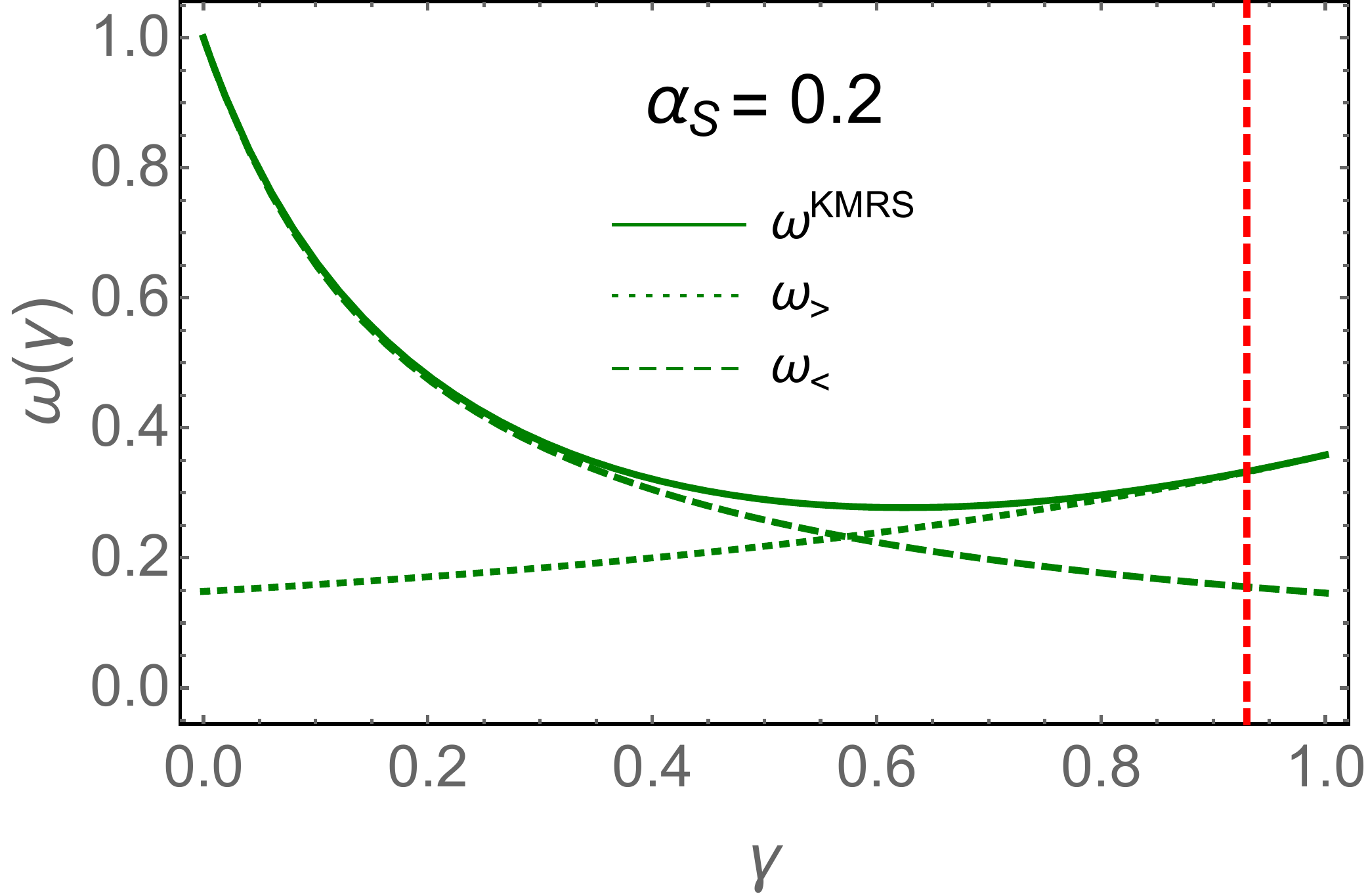}\\
      \fig{om}-a & ~~~~~~~~~~~&\fig{om}-b\\
 \end{tabular}
         \caption{The eigenvalues $\omega^{\rm KMRS}$ versus $\gamma$. 
The vertical dashed lines show the value of $\bar{\gamma} = 1 -
 \gamma_{cr}$, which is given by \eq{GACR},  for different values
 of $\bas$. $\omega_{>}$ is given by \eq{OMNLOKMRS1}, while 
 $\omega_{<}$ is  solution to         
the following equation:   $
 \omega_{<}\,\,=\,\,\bas \Lb 1 \,-\,\omega_{<}\Rb \Bigg\{ \frac{1}{\gamma} \,\,-\,\,\omega_{<}\Bigg\}$ which we will discuss below in section   IV-A.}    
\label{om}
\end{figure}

All other terms in  \eq{KMRSOM} vanish at $\gamma = 1$. 
 Solving \eq{OMNLOKMRS} we obtain

\beq \label{OMNLOKMRS1}
\omega\,\,\equiv \omega_{>}\,\,=\,\,\h\Lb-\Lb 1 \, -\, \gamma\,+\,\bas\Rb\,+\,\sqrt{4\,\bas \,+\,\Lb 1\, -\, \gamma\,+\,\bas\Rb^2}\Rb
\eeq
\eq{OMNLOKMRS1} gives a good description of  the eigenvalues
 of \eq{KMRSOM} $\omega^{\rm KMRS}$ for $\gamma > 1 - 
 \gamma_{cr}=\bar{\gamma}$ (see \fig{om}). This is the
 reason  that we  denote this eigenvalue as $\omega_{>} $.

\subsection{Saturation and geometric scaling behaviour of the scattering
 amplitude in the NLO}


It is well known\cite{KOLEB,GLR,BALE,LETU, IIML,MUT}, that one does
 not need to know the precise structure of the non-linear corrections
 for finding the saturation momentum, as well as for discussing the 
behaviour of the scattering amplitude in the vicinity of the saturation
 scale.  We only need  to find the solution of the linear BFKL equation,
 which is  a wave package that satisfies the condition, that phase and
 group velocities are equal\cite{GLR}. This solution determines the new
 dimensional scale of the problem: the saturation momentum  $Q_s$.  The
 equation takes the form
\beq \label{EQQS}
 v_{\rm group} =  \,\frac{d \omega\Lb \gamma\Rb}{d \gamma}\,=\,v_{\rm
 phase}\,=\,\frac{\omega\Lb \gamma\Rb}{  \gamma}
 \eeq
For the eigenvalues of the BFKL equation in the NLO given by \eq{OMNLO},
 the solution to \eq{EQQS} takes the form:
\beq \label{GACR}
\bar{\gamma}\,\,=\,\,1\,-\,\gamma_{cr}\,\,=\,\,\h + 2 \bas
\eeq
The equation 
\bea \label{PHASE}
&&\frac{\omega\Lb \bar{\gamma}\Rb}{ \bar{\gamma}}\,Y\,\,-\,\,\xi_s\,\,=\,\,\frac{4 \bas}{1 \,+\,4 \bas}\,Y\,\,-\,\,\xi_s\,\,=\,\,0;~~~\ln\Big( Q^2_s/Q^2_0\Lb Y=Y_0\Rb\Big)\,=\,\frac{4 \bas}{1 \,+\,4 \bas}\, \Lb  Y  - Y_0\Rb;\nn\\
&&
Q^2_s\,=\,Q^2_s\Lb Y = Y_0\Rb\exp\Lb\frac{4 \bas}{1 + 4 \bas} \Lb  Y  - Y_0\Rb \Rb
\eea
determines the saturation momentum $\xi_s\,\,=\,\,-\ln\Lb r^2_s\,Q^2_s\Lb Y =
 Y_0\Rb\Rb\,\,=\,\, \ln\Lb  Q^2_s\Lb Y\Rb/Q^2_s\Lb Y_0\Rb\Rb$\footnote
{Actually $Q^2_s\Lb Y=Y_0\Rb$ depends on other variables as well (see
 \eq{XI1}), but we will omit these variable in the further presentation.}.

\begin{figure}
\centering
      \includegraphics[width=9cm]{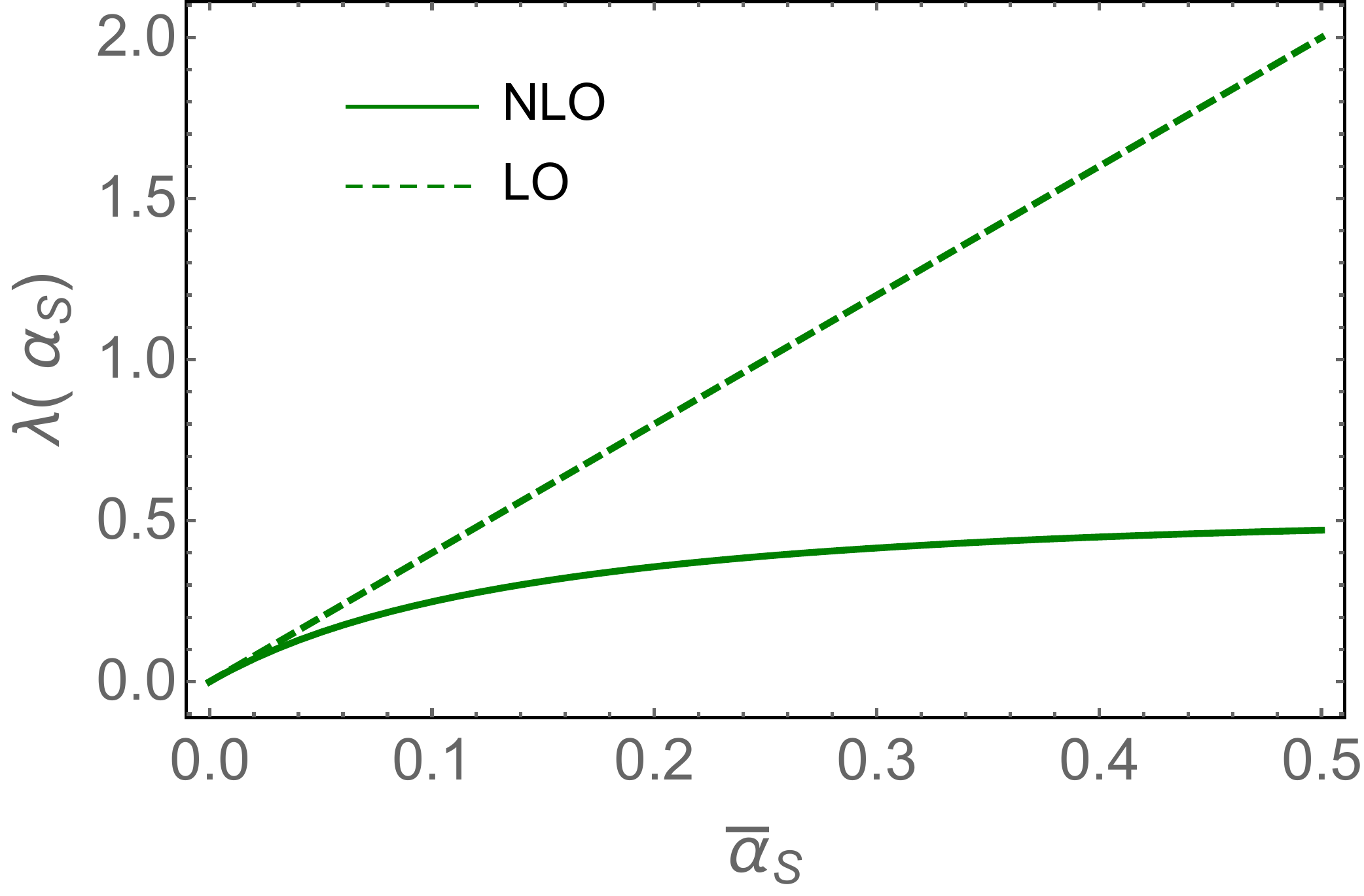}
   \caption{ $\lambda\Lb \bas\Rb$, which determines the energy dependence
 of the saturation scale $Q^2_s\,\, \propto \,\,\Lb \frac{1}{x}\Rb^{
\lambda\Lb\bas\Rb} $, versus $\bas$.}    
\label{la}
\end{figure}

For \eq{OMNLOKMRS1} it turns out that 
\beq \label{PHASE1}
\bar{\gamma}\,\,=\,\,1\,-\,\gamma_{cr} \,\,=\,\, \frac{1 + 6 \bas + \bas^2)}{2 (\bas +1)};~~~~~
\,\,\ln\Lb  Q^2_s\Lb Y\Rb/Q^2_s\Lb Y_0\Rb\Rb=\,\frac{4 \bas}{1 \,+\,6 \bas\,+\,\bas^2}\,\Lb Y - Y_0\Rb\,\,\equiv\,\,\lambda\,\Lb Y - Y_0\Rb\,;
\eeq
From \fig{la} one can see that \eq{PHASE1} leads to the value of $\lambda
 \approx 0.25$, which describes the experimental data at $\bas
 \,\,\approx\,\, 0.2$.

In the vicinity of the saturation scale where $r^2 Q^2_s\Lb Y\Rb \,=\,\tau
 \to 1$, the scattering amplitude has the following form\cite{IIML,MUT}:

\beq \label{VQS}
N\Lb r^2, Y\Rb\,\,=\,\,N_0 \Lb r^2\,Q^2_s\Rb^{1 - \gamma_{cr}}\,\,=\,\,N_0\,\,\tau^{1 - \gamma_{cr}}\eeq
This equation gives  the initial conditions  for the scattering amplitude
 in the saturation domain. These conditions have  the forms:
\beq \label{IC}
N\Lb r^2, Y\Rb\,\,=\,\,N\Lb \tau=1\Rb\,\,=\,\,N_0;~~~~~~~
\frac{d \ln \Lb N\Lb r^2, Y\Rb\Rb}{d \xi}|_{\tau=1} \,\,=\,\,1 - \gamma_{cr};
\eeq

\subsection{Double log approximation (DLA)   in the re-summed NLO approximation}

\subsubsection{DLA in the LO approximation}

   In the leading order BFKL equation, the DLA stems from the eigenfunction
   
   \beq \label{DLALO1}
   \omega\Lb \gamma\Rb \,\,=\,\,\frac{\bas}{ 1 - \gamma}
   \eeq
   In the coordinate representation (see  the linear part of \eq{BK})
 the DLA comes from the distances  $x_{12} \sim x_{02}\,\,\gg\,\,x_{01}$.
 For such distances the BFKL equation has the form
   \bea \label{DLALO2}
&&\frac{\partial}{\partial Y}N\Lb \vec{x}_{10}, \vec{b} ,  Y \Rb = \nn\\
&&\bas\!\! \int \frac{d^2 \vec{x}_2}{2\,\pi}\,K\Lb \vec{x}_{02}, \vec{x}_{12}; \vec{x}_{10}\Rb \Bigg(N\Lb \vec{x}_{12},\vec{b} - \h \vec{x}_{20}, Y\Rb + 
N\Lb \vec{x}_{20},\vec{b} - \h \vec{x}_{12}, Y\Rb - N\Lb \vec{x}_{10},\vec{b},Y \Rb\Bigg)\nn\\   
 && \xrightarrow{ x_{12} \sim x_{02}\,\,\gg\,\,x_{01}}\,\,\bas\,x^2_{01}\,\int_{x^2_{01}} \frac{d x^2_{12}}{x^4_{12} }  N\Lb \vec{x}_{12},\vec{b} , Y\Rb   
 \eea
 In the derivation of \eq{DLALO2} we used \eq{KERLO} for the kernel and
 assumed that $b \,\gg\,x_{12}$. Note, that the gluon reggeization term
 ($ N\Lb \vec{x}_{10},\vec{b},Y \Rb$)  in \eq{DLALO2} does not contribute
 in the DLA.
 
 Introducing $\tilde{N}\Lb x_{ij}\Rb \,=\,N(x_{ij})/x^2_{ij}$ and  changing
 $\xi \,\to \xi' = - \xi$ we obtain \eq{DLALO2} in the traditional form:
 \beq \label{DLALO3}
 \frac{\partial}{\partial Y}\tilde{N}\Lb \xi',   Y; b \Rb =
 \bas \int^{\xi'} d \xi'' \,\tilde{N}\Lb \xi'', Y; b\Rb\,;~~~~~~~~~
 \frac{\partial^2}{\partial Y\,\partial\,\xi'}\tilde{N}\Lb \xi',  
 Y; b  \Rb\,\,=\,\, \bas \, \tilde{N}\Lb \xi',Y; b\Rb\,.\eeq
 
 \eq{DLALO3} follows naturally from the general solution to
 the BFKL equation\footnote{The dependance on $b$ appears in
 the definition of $\xi$ (see \eq{EIGENF}) and in  the initial
 condition (function $\tilde{n}_{in}\Lb \gamma\Rb$). In the following
 we omit  $b$ dependance and hope, that it will not cause any difficulties.}:
 
 \beq \label{GENSOL}
 \tilde{N}\Lb \xi,Y; b\Rb\,\,=\,\,\int^{\epsilon + i \infty}_{\epsilon - i \infty} \frac{ d \omega}{ 2\,\pi\,i} \, \,\int^{\epsilon + i \infty}_{\epsilon - i \infty} \frac{ d \gamma}{ 2\,\pi\,i}\frac{1}{ \omega\,\,-\,\,\omega\Lb \gamma\Rb}\,e^{ \omega\,Y \, - \,\Lb 1 - \gamma\Rb \xi} \tilde{n}_{in}\Lb \gamma\Rb
 \eeq
 using the eigenvalue of \eq{DLALO1}.  Indeed, plugging \eq{DLALO1} into
 \eq{GENSOL}  and taking the integral over $\gamma$, closing the contour
 of integration we obtain
 \beq \label{DLALO4}
 \tilde{N}\Lb \xi,Y; b\Rb\,\,= \,\,\int^{\epsilon + i \infty}_{\epsilon - i \infty} \frac{ d \omega}{ 2\,\pi\,i}\,e^{ \omega\,Y \, - \, \frac{\bas}{\omega}\,\xi}\,\,\tilde{n}_{in}\Lb \omega\Rb 
 \,\,=\,\, \sum^{\infty}_{n=0}\frac{ \xi'^n}{n!}\int^{\epsilon + i \infty}_{\epsilon - i \infty} \frac{ d \omega}{ 2\,\pi\,i}\,\Lb  \frac{\bas}{\omega} \Rb^n\,e^{ \omega\,Y}\,\,\tilde{n}_{in}\Lb \omega\Rb \eeq
   Taking the integral over $\omega$ we obtain the solution which is a
 series, with a general term which is proportional $\Lb \bas \,Y\, 
\xi'\Rb^n$.
 The particular form of the solution is determined by the function
 $ \tilde{n}_{in}\Lb \omega\Rb $ which we can find from the initial
 condition at  $Y=0$.

\subsubsection{DLA in the NLO for BFKL equation}

 
 We wish to find the DLA  in the NLO , using \eq{GENSOL}  in which we substitute  the solution\footnote{\eq{DLANLO1} was  derived from \eq{KMRSOM} in Ref.\cite{ASV} and has been discussed in Ref.\cite{DIMST}.} to
 \eq{GANLOS} with respect to $\gamma$:
  \beq \label{DLANLO1}
 1 \,-\,\gamma\,\,\,=\,\,\,\frac{\bas}{\omega} \,\,-\,\,\omega
 \eeq
 Plugging \eq{DLANLO1} into \eq{GENSOL} we obtain
  \beq \label{DLANLO2}
 \tilde{N}\Lb \xi,Y; b\Rb\,\,= \,\,\int^{\epsilon +
 i \infty}_{\epsilon - i \infty} \frac{ d \omega}{ 2\,\pi\,i}\,e^{
 \omega\,\Lb  Y \,\,+\,\, \xi\Rb\, - \, \frac{\bas}{\omega}\,\xi}\,
\,\tilde{n}_{in}\Lb \omega\Rb 
 \,\,=\,\, \sum^{\infty}_{n=0}\frac{ \xi'^n}{n!}\int^{\epsilon +
 i \infty}_{\epsilon - i \infty} \frac{ d \omega}{ 2\,\pi\,i}\,\Lb 
 \frac{\bas}{\omega} \Rb^n\,e^{ \omega\,\eta}\,\,\tilde{n}_{in}\Lb
 \omega\Rb \eeq 
 
where we have  introduced new variables: $\eta \,\,=\,\,Y \,\,-\,\,\xi'$
 and $\xi' \,=\,-\xi$. From \eq{DLANLO2} one can see  that the solution
 is a series with the general term  $\propto\,\Lb \bas \,\eta\,\xi'\Rb^n$,
 whose exact form is determined by the initial condition at $Y = 0$.
 
 The solution of \eq{DLANLO2} can be re-written in more  economical form:

  \beq \label{DLANLO3}
 \frac{\partial}{\partial \,\eta}\tilde{N}\Lb \xi', \eta; b \Rb =
 \bas \int^{\xi'} d \xi'' \,\tilde{N}\Lb \xi'',\eta; b\Rb\,;~~~~~~~~~
 \frac{\partial^2}{\partial \,\eta\,\partial\,\xi'}\tilde{N}\Lb \xi',
 \eta; b \Rb\,\,=\,\, \bas \, \tilde{N}\Lb \xi',\eta; b\Rb\,,\eeq

 Therefore, we see  that the only difference of the DLA in the NLO, stems
 from a  new variable $\eta,$ which replaces $Y \,=\,\ln (1/x)$. The 
physical
 meaning of this replacement has been examined in detail in Ref.\cite{DIMST}.
  The  point  of it is the fact, that the partons (gluons) in the 
wave function
 of the fast hadrons are ordered  in accord of the lifetimes of the partons,
  which are proportional $t_i \,\sim x_i  \,P/k^2_{i,T}$, where $x_i$ 
denotes the
 fraction of the longitudinal momentum of the fast hadron, with the moment
 $P$, and $k_{i,T}$ is the transverse momentum of the $i$-th parton. The
 ordering that leads to the logs in $\eta$, has the form $ t_1\,
 \gg\,t_2\,\gg\,\dots\,t_{i-1}\,\gg\,t_i \gg\,\,\dots\,t_{n}$.

 It is instructive to note, that the form of the kernel and the
 differential \eq{DLANLO3}
  are quite different,  from the ones that has been discussed in
 Refs.\cite{DIMST,CLM}. Indeed, in Ref.\cite{DIMST} it is suggested
 that the linear equation has the form:
  \beq \label{DLANLO4}
\frac{d N\Lb r, Y; b\Rb}{d   Y}\,
\,=\,\,\bas \int \frac{d r'^2 \, r^2}{ r'^4}\,\,\frac{J_1\Lb 2  \sqrt{\bas \rho^2}\Rb}{\sqrt{\bas \rho^2}}\, N\Lb r', Y; b \Rb
\eeq  
  where $\rho\,\equiv\,\sqrt{L_{x_{02}, x_{01}}L_{x_{12}, x_{01}}}$ and
 $L_{ x_{i 2}, x_{01}} \equiv \ln(x_{i 2}^2/x_{01}^2)$. 
  
  The solution to \eq{DLANLO4} has the following form for $\eta =
 Y\,-\,\xi'\,>\,0$ \cite{DIMST,CLM}
 
\beq \label{DLANLO5}
N\Lb\xi',Y\Rb\,\,\,=\,\,e^{-\xi'}\,\,\phi_{in}\Lb 0\Rb\,\,Y\,\Lb  \displaystyle{  \dfrac{I_{1}\Lb 2\,\sqrt{\bas\, \xi' \,\eta }\Rb}{ \sqrt{\bas\, \xi' \,\eta}  } }\Rb\eeq 
  
  Assuming that solution to  \eq{DLANLO3}  depends on one variable $\zeta
 \,=\,2\,\sqrt{\,\bas\,\xi'\,\eta}$ we obtain that this equation takes the
 form:
  \beq \label{DLANLO6}   
  \frac{d^2\,\widetilde{N}\Lb \zeta\Rb}{d\,\zeta^2}\,\,+\,\,\frac{1}{\zeta} \frac{d\,\widetilde{N}\Lb \zeta\Rb}{d\,\zeta}  \,\,=\,\,\,\widetilde{N}\Lb \zeta\Rb  
  \eeq
  which has the  solution $\widetilde{N}\Lb \zeta\Rb\,\,=\,\,C_1\,I_0\Lb
 \zeta\Rb$ (see formula {\bf 8.494(1)} in Ref.\cite{RY}).
  
Summing double logs and comparing these two solutions, one can see that 
both
 are function of $\bas\,\eta\,\xi'$. Both solutions at $\bas\,\eta\,\xi'\,
\,\gg\,1$ have the asymptotic behaviour
  \beq \label{DLANLO7}
N\Lb\xi',Y\Rb\,\,\,\xrightarrow{\zeta\,\gg\,1}\,\,H\Lb Y, \xi\Rb\,e^{2
 \,\sqrt{ \bas\,\eta\,\xi'}\,\,-\,\,\,\xi'}\,\,\eeq     

The function $H$  depends on $Y$ and $\xi'$  logarithmically. 

The expression  for the saturation momentum stems  from the equation
\beq \label{DLANLO08}
2 \,\sqrt{ \bas\,\eta\,\xi'}\,\,-\,\,\,\xi'\,\,=\,\,0
\eeq
which gives the condition when $N\,\sim\,H\Lb Y, \xi\Rb \,\,\sim\,{\rm
 Const}$, since $H$ is slowly changing function.

It is easy to see that \eq{DLANLO08} leads to the saturation momentum of
 \eq{PHASE}.

However, to  understand better  the difference between these two
 equations, we would like to re-write \eq{DLANLO4} in the form
 of \eq{DLANLO3}. The general solution to \eq{DLANLO4} has the
 following form:
\beq \label{DLANLO90}
 \widetilde{N}\Lb \xi',  Y \Rb\,\,\,=\,\,\,\int^{\epsilon + i \infty}_{\epsilon 
- i \infty}\frac{d \gamma'}{2\,\pi\,i} e^{ \h\Lb- \gamma'\,+\,\sqrt{4\,\bas \,+\,\gamma'^2}\Rb\,Y \,\,+\,\,\gamma'\,\xi'} \,\phi_{\rm in}\Lb \gamma', R\Rb
 \eeq
where we use \eq{OMNLO} for $\omega(\gamma)$, and replace $1\,-\,
\gamma$ by $\gamma'$. The function $\phi_{\rm in}\Lb \gamma', R\Rb$
 is determined by the initial condition at $Y=0$  and  $R$ denotes the 
size
 of the target dipole. 

Using \eq{DLANLO90} we  see that $ \widetilde{N}\Lb \xi',  Y \Rb$
 satisfies the following equation:
\bea \label{DLANLO9}
&&\frac{\partial^2\, \widetilde{N}\Lb \xi',  Y \Rb}{\partial Y\,\partial\,\xi'}\,\,+\,\,\frac{\partial^2\, \widetilde{N}\Lb \xi',  Y \Rb}{\partial Y^2} \,\,=\,\, \int^{\epsilon + i \infty}_{\epsilon 
- i \infty}\frac{d \gamma'}{2\,\pi\,i} \Big( \omega\Lb \gamma'\Rb\,\gamma' \,\,+\,\,
\omega^2\Lb \gamma'\Rb\Big)
e^{ \h\Lb- \gamma'\,+\,\sqrt{4\,\bas \,+\,\gamma'^2}\Rb\,Y \,\,+\,\,\gamma'\,\xi'} \,\phi_{\rm in}\Lb \gamma', R\Rb \nn\\
&&=\,\, \int^{\epsilon + i \infty}_{\epsilon 
- i \infty}\frac{d \gamma'}{2\,\pi\,i} \Bigg\{\h\Big(- \gamma'\,+\,\sqrt{\gamma'^2 \,+\,4\,\bas}\Big) \,\gamma' \,\,+\,\,\frac{1}{4}\Big(- \gamma'\,+\,\sqrt{\gamma'^2 \,+\,4\,\bas}\Big)^2\Bigg\}\,
e^{ \h\Lb- \gamma'\,+\,\sqrt{4\,\bas \,+\,\gamma'^2}\Rb\,Y \,\,+\,\,\gamma'\,\xi'} \,\phi_{\rm in}\Lb \gamma', R\Rb\,\nn\\
&&\,\,=\,\,\bas \, \widetilde{N}\Lb \xi',  Y \Rb
\eea

Assuming that $ \widetilde{N}\Lb \xi',  Y \Rb\,\,=\,\, \widetilde{N}\Lb
 \xi',\eta \equiv   Y\,-\,\xi' \Rb$    \eq{DLANLO9}, can be reduced to
 \eq{DLANLO3}. Indeed, one can see that the solution of \eq{DLANLO2}
 satisfies both these equations.


\subsubsection{Non-linear evolution in   LO  DLA}

\paragraph{Gluon reggeization:}

The linear evolution equation  in the leading order DLA 
is given by \eq{DLALO3}.  To include    non-linear corrections,
 we need to consider the general BK equation (see \eq{BK}),  with the
 kernel in the DLA approximation. The first question which arises, is
  how to treat the reggeization term in \eq{DLALO2}, which is neglected
 in the DLA. Indeed, this term does not contribute in the DLA and 
including it,  is a particular way to make estimates beyond those
 of the DLA. We believe that it is necessary to do this, to provide
 the correct behaviour of the scattering amplitude which should approach
  1 ($N \,\to\,1$) for large $Y$.  Bearing this in mind we need to 
 preserve the gluon  reggeization term in \eq{DLALO2}, which has the
 form

 \bea \label{DLALONL1}
&&\frac{\partial}{\partial Y}N\Lb \vec{x}_{10}, \vec{b} ,  Y; R \Rb = \nn\\
&&\bas\!\! \int \frac{d^2 \vec{x}_2}{2\,\pi}\,K\Lb \vec{x}_{02}, \vec{x}_{12}; \vec{x}_{10}\Rb \Bigg\{N\Lb \vec{x}_{12},\vec{b} - \h \vec{x}_{20}, Y; R\Rb + 
N\Lb \vec{x}_{20},\vec{b} - \h \vec{x}_{12}, Y; R\Rb - N\Lb \vec{x}_{10},\vec{b},Y \Rb\Bigg\}\nn\\   
 && \xrightarrow{ x_{12} \sim x_{02}\,\,\gg\,\,x_{01}}\,\,\h\,\bas\,x^2_{01}\,\int_{x^2_{01}} \frac{d x^2_{12}}{x^4_{12} }  \Bigg( 2\,N\Lb \vec{x}_{12},\vec{b} , Y; R\Rb \,\,-\,\, N\Lb \vec{x}_{01},\vec{b} , Y; R\Rb \Bigg)
 \eea
 Using $\tilde{N}\,\,=\,\,R^2\,N/x^2_{01}$ and $\xi' =
 \ln\Lb R^2/x^2_{01}\Rb$\footnote{In definition of $\xi'$ we
 assumed that $b \,\ll\,R$. For $b \,\gg\,R $, $\xi' \,\,=\,
\,\ln\Lb b^2/x^2_{01}\Rb$.} we can re-write \eq{DLALONL1}
 in the form:
 
 \beq \label{DLALONL2}
 \frac{\partial^2}{\partial Y\,\partial\,\xi'}\tilde{N}\Lb \xi',Y \Rb \,\,=\,\,\bas \,\tilde{N}\Lb \xi',Y \Rb \,\,-\,\,\h\,\bas
  \frac{\partial}{\partial\,\xi'}\tilde{N}\Lb \xi',Y \Rb 
  \eeq

 This equation corresponds to the eigenvalue $\omega(\gamma)$
 for the amplitude  $\tilde{N}\Lb \xi',Y\Rb$:
  \beq \label{DLALONL3} 
  \omega\Lb \gamma\Rb\,\,=\,\,\frac{\bas}{\gamma} \,\,-\,\,\h\,\bas
  \eeq

  For the dipole amplitude, $N\Lb \xi',Y\Rb$, \eq{DLALONL1} takes
 the following form:

 \beq \label{DLALONL20}
 \frac{\partial^2}{\partial Y\,\partial\,\xi'} N \Lb \xi',Y \Rb \,\,+\,\, \frac{\partial}{\partial Y} N \Lb \xi',Y \Rb\,\,=\,\,\h\,\bas \,N\Lb \xi',Y \Rb \,\,-\,\,\h\,\bas
  \frac{\partial}{\partial\,\xi'} N\Lb \xi',Y \Rb 
  \eeq  
  which leads to the eigenvalue $\omega(\gamma)$ :
   \beq \label{DLALONL21}  
  \omega\Lb \gamma\Rb\,\,=\,\,\h\,\bas\,\frac{1 \,+\,\gamma}{1\,-\,\gamma}
  \eeq
  
    Applying the procedure, that has been discussed in subsection III-C  we
 obtain
  
  \beq \label{DLALONL4}  
   \ln\Lb Q^2_s\Lb Y\Rb/Q^2_s\Lb Y= 0\Rb\Rb\,\,=\,\,\h \bas \Lb 3 + 2 \sqrt{2}\Rb\,Y\,\,=\,\,\lambda\,Y; ~~~~~\mbox{\rm and}~~~\bar{\gamma}\,\,=\,\,\,\sqrt{2}\,-\,1
  \eeq
 We would like to mention that the value of $\lambda $ turns 
out
 to be  30\% less that the LO  DLA value, $\lambda = 4\,\bas$.
 
 \paragraph{Non-linear equation:}

Finally, we need to re-write the general BK equation (see \eq{BK}) and
 account for the non-linear term.
Using the BFKL kernel in the DLA we obtain
 
     \bea \label{DLALONL5}
 \frac{\partial}{\partial \,Y}\tilde{N}\Lb \xi',Y \Rb &=& \bas \int^{\xi'} d \xi'' \,\Bigg\{ \tilde{N}\Lb \xi'', Y\Rb\,\,-\,\,\h\,e^{-\xi''} \tilde{N}^2\Lb \xi'',Y\Rb \Bigg\}\,\,-\,\,\h\, \bas \tilde{N}\Lb \xi',Y\Rb;\nn\\
 \frac{\partial^2}{\partial \,Y\,\partial\,\xi'}\tilde{N}\Lb \xi', \ Y \Rb\,\,&=&\,\, \bas \,\Bigg\{ \tilde{N}\Lb \xi',Y \Rb\,\,-\,\,\h\,e^{-\xi'} \tilde{N}^2\Lb \xi',Y\Rb \Bigg\}
 \,\,-\,\,\h\, \bas\frac{\partial}{\partial \,\xi'}\tilde{N}\Lb \xi',Y\Rb 
 .\eea

    For the dipole amplitude $N\,=\,e^{ - \xi'} \tilde{N}$,  \eq{DLALONL5}
 has the following form
    
    \beq \label{DLALONL60}
         \frac{\partial^2}{\partial \,Y\,\partial\,\xi'} \,N\Lb \xi', Y \Rb\,\,+\,\, \frac{\partial}{\partial \,Y}\,N\Lb \xi', Y  \Rb=\,\, \h\, \bas \,\Bigg\{N\Lb \xi',Y; \vec{b}\Rb\,\,-\,\, N^2\Lb \xi', Y \Rb \Bigg\}\,\,-\,\,\h\,\bas \frac{\partial}{\partial \,\xi'}\,N\Lb \xi', Y \Rb    \eeq    
     
     \eq{DLALONL60} has the traveling wave solution,  which is a
 function of one variable $ z\,\,=\,\,\alpha\,Y\,\,+\,\,\beta\,\xi'$.
 Indeed \eq{DLALONL60} takes the form
    
    \beq \label{DLALONL6}  
     \alpha\,\beta\,  \frac{d^2}{d z^2} \,N\Lb z \Rb\,\,+\,\,\Lb \alpha \,\,+\,\,\h\,\bas\,\beta\Rb\frac{d}{d \,z}\,N\Lb z  \Rb=\,\,\h\, \bas \,\Bigg\{ N\Lb z\Rb\,\,-\,\, N^2\Lb z\Rb \Bigg\}
     \eeq    
    
 From \eq{DLALONL3}  the natural choice for  the parameters $\alpha$
 and $\beta$ is $\alpha\,\,=\,\,\h \,\bas\,\Lb 3 + 2\sqrt{2}\Rb$ and 
 $\beta\,=\,-1$.
 
 We postpone the discussion of the nonlinear equation to the next section. 
 For small $N\,\ll\,1$,  we can neglect the non-linear term,  and
 \eq{DLALONL6} leads to the scattering amplitude $N$ which is equal to
 \beq \label{DLALONL7}  
   \,N\Lb z \Rb\,\,=\,\,N_0\,\exp\Lb \Lb \sqrt{2} \,-\,1 \Rb\,z\Rb\,\,=
\,\,N_0\Lb r^2\,Q^2_s\Lb Y\Rb\Rb^{ \sqrt{2} \,-\,1}     \eeq   
this  coincides with the general form of  \eq{VQS} where $\lambda $
 and $\bar{\gamma}$   are determined by \eq{DLALONL4}.

 \subsubsection{Non-linear evolution in NLO  DLA:}
 
\paragraph{Equation:}


As we have discussed in section III-D-2,  the difference between LO DLA
 and NLO DLA (see \eq{DLANLO1}) stems only from the new energy variable:
 $\eta \,=\,Y\,\,-\,\,\xi'$. Therefore, the non-linear equations  have
 the form:

     \bea \label{DLANL1}
 \frac{\partial}{\partial \,\eta}\tilde{N}\Lb \xi',   \eta \Rb &=& \bas \int^{\xi'} d \xi'' \,\Bigg\{ \tilde{N}\Lb \xi'',\eta\Rb\,\,-\,\,\h\,e^{-\xi''} \tilde{N}^2\Lb \xi'',\eta\Rb \Bigg\}\,\,-\,\,\h\, \bas \tilde{N}\Lb \xi',\eta\Rb;\nn\\
 \frac{\partial^2}{\partial \,\eta\,\partial\,\xi'}\tilde{N}\Lb \xi', \eta \Rb\,\,&=&\,\, \bas \,\Bigg\{ \tilde{N}\Lb \xi',\eta \Rb\,\,-\,\,\h\,e^{-\xi'} \tilde{N}^2\Lb \xi',\eta\Rb \Bigg\}
 \,\,-\,\,\h\, \bas\frac{\partial}{\partial \,\xi'}\tilde{N}\Lb \xi',\eta;\Rb 
 ;\eea 
   
   The last term in \eq{DLANL1} comes from the gluon reggeization (the
 third term in \eq{BK}). It does not contribute  to the DLA approximation,
  but has to be taken into account together with the non-linear corrections
, since it provides the correct asymptotic behaviour of the scattering 
amplitude
 $N\,\to\,1$ at $Y \,\,\gg\,\,1$.

   The linear equation  for the amplitude $\tilde{N}\Lb  \xi',  Y\Rb$ ,
 whose origin is \eq{DLANL1} neglecting the non-linear term, can be
 written, using \eq{DLALONL2},  as the following equation for the
 eigenvalue $\omega(\gamma')$:
      \beq \label{DLANL100}   
    \omega(\gamma')\Big( \gamma' \,\,+\,\,\omega(\gamma')\Big) \,\,=\,\,\bas\,\,-\,\,\h\,\bas \,\gamma'
    \eeq    
    
     \eq{DLANL100} differs from \eq{DLANLO1}, by the last
 term in the r.h.s. of this equation, which takes into account the gluon
 reggeization contribution.
    
The solution which has the following form:
  
    \beq \label{DLANLO101}
 \widetilde{N}\Lb \xi',  Y \Rb\,\,\,=\,\,\,\int^{\epsilon + i \infty}_{\epsilon 
- i \infty}\frac{d \gamma}{2\,\pi\,i} e^{ \omega(\gamma')\,Y \,\,+\,\,\gamma'\,\xi'} \,\phi_{\rm in}\Lb \gamma', R\Rb
 \eeq    
    where $\omega(\gamma')$ is taken from \eq{DLANL100}.
    
\begin{figure}
\begin{tabular}{c  c  c}
      \includegraphics[width=7cm]{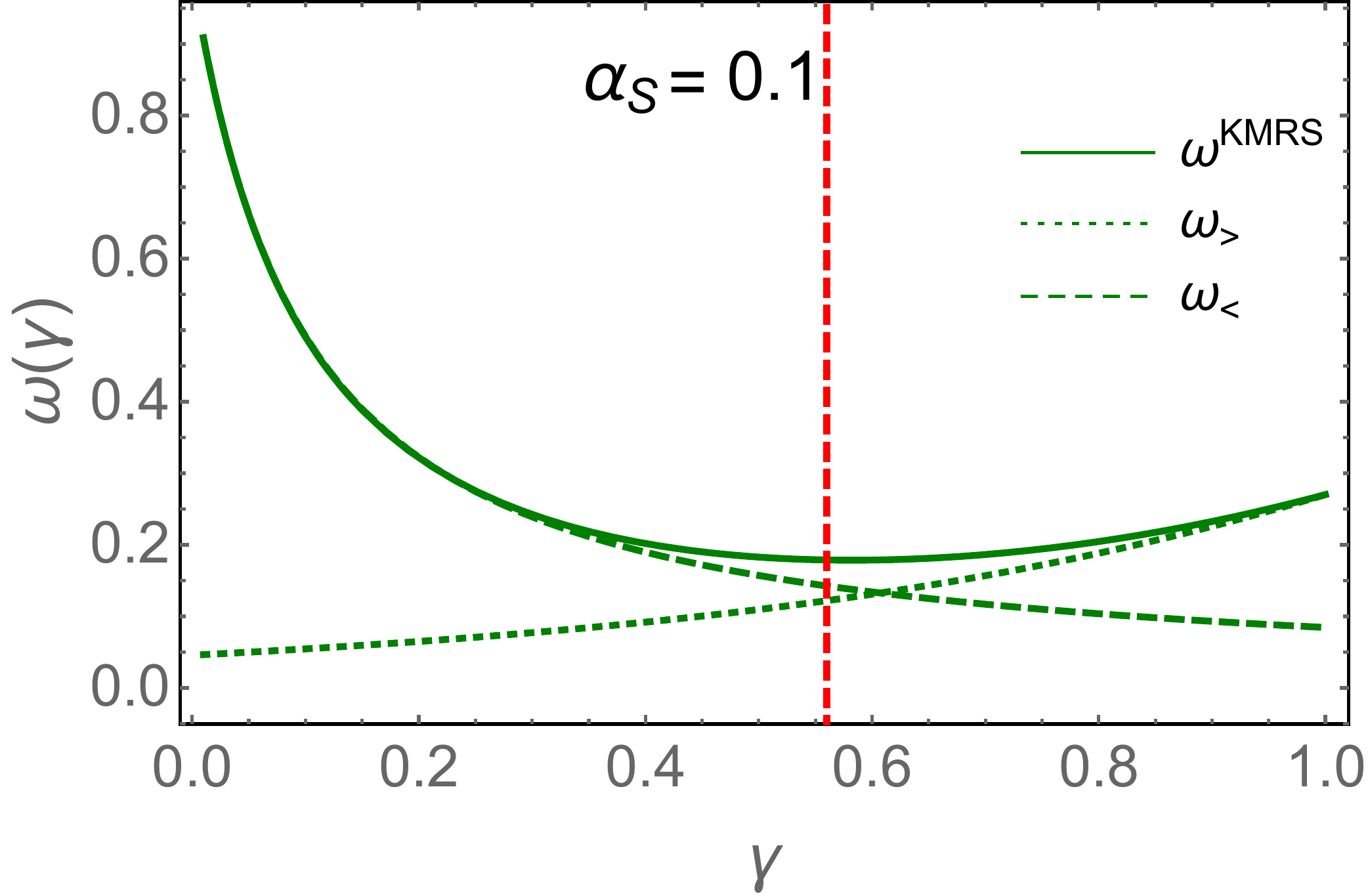}&  & \includegraphics[width=7cm]{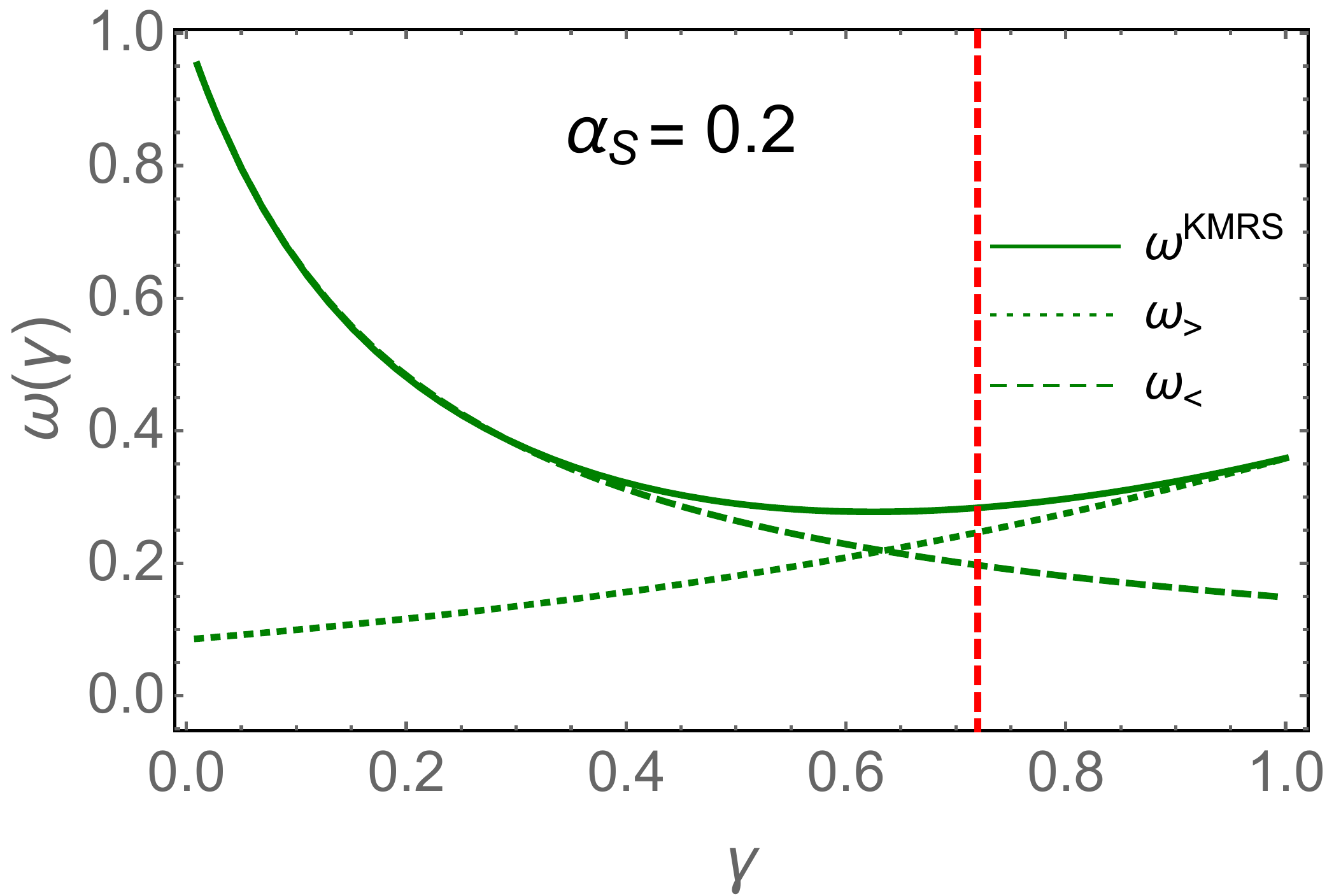}\\
      \fig{omi}-a & ~~~~~~~~~~~&\fig{omi}-b\\
 \end{tabular}
         \caption{The eigenvalues $\omega$ versus $\gamma$. The vertical
 dashed lines show the value of $\bar{\gamma} = 1 - \gamma_{cr}$ for
 different values of $\bas$.  $\omega_{>}\,\equiv\,\omega_{m}$ is given
 by \eq{DLANL120}. $\omega_{<}$ is the same as in \fig{om}, which we 
 discuss below in section   IV-A . $\omega^{\rm KMRS}$ is given by
 \eq{KMRSOM}.}      
\label{omi}
\end{figure}

     In \fig{omi} we plot $\omega_{m} $, which is the solution to the 
equation:
      \beq \label{DLANL11}   
    \omega_{m}\Big( \gamma \,\,+\,\,\omega_{m}\Big) \,\,=\,\Lb 1\,\,-\,\,\omega_{m}\Rb\,\Lb\bas\,\,-\,\,\h\,\bas \,\gamma\Rb
    \eeq       
   This equation is the generalization of  \eq{DLANL100} in where we
 took into account,  energy conservation.
        
  Note, that this equation introduces   conservation of energy into
 \eq{DLANL100}.  The solution to  \eq{DLANL11}  has the form:
     \beq \label{DLANL120}       
   \omega_{m}\Lb \gamma\Rb\,\,=\,\,\frac{1}{4} \left(\bas \gamma +\sqrt{(\bas  \gamma -2 \bas -2 \gamma )^2+8 (2 \bas -\bas  \gamma )}-2 \bas -2 \gamma \right)  
     \eeq   


    For the dipole amplitude $N\,=\,e^{ - \xi'} \tilde{N}$ \eq{DLANL1} has
 the following form
    
    \beq \label{DLANL2}
     \frac{\partial^2}{\partial \,\eta\,\partial\,\xi'} \,N\Lb \xi', \eta \Rb\,\,+\,\, \frac{\partial}{\partial \,\eta}\,N\Lb \xi',  \eta \Rb=\,\, \h\, \bas \,\Bigg\{N\Lb \xi', \eta \Rb\,\,-\,\, N^2\Lb \xi',\eta\Rb \Bigg\}\,\,-\,\,\h\,\bas \frac{\partial}{\partial \,\xi'}\,N\Lb \xi', \eta\Rb    \eeq    
     
For  \eq{DLANL2}, as well as for \eq{DLALONL60}, we can find a  solution,
  which is the function of one variable $ z\,\,=\,\,\alpha\,\eta\,\,+\,
\,\beta\,\xi'$. Indeed, the equation  \eq{DLANL2} takes the form
    
    \beq \label{DLANL3}
   \alpha\,\beta\,  \frac{d^2}{d z^2} \,N\Lb z\Rb\,\,+\,\,\Lb \alpha \,\,+\,\,\h\,\,\bas\,\beta\Rb\frac{d}{d \,z}\,N\Lb z \Rb=\,\,\h\, \bas \,\Bigg\{ N\Lb z\Rb\,\,-\,\, N^2\Lb z \Rb \Bigg\}
     \eeq    
    
    As we have seen in  the discussion of \eq{DLALONL60},  the natural 
choice
 of  parameters $\alpha$ and $\beta$ is
    $\alpha\,=\,\,\h\,\bas\,\Lb 3 + 2\sqrt{2}\Rb\,\equiv\,a\,\bas$
 and $\beta \,=\,-1$.
   \paragraph{Solution: generalities:}     
    We have not found the explicit solution to \eq{DLANL3}. This is           
the 
reason for  discussing   the general features of the solution in this  
subsection, based on the phase portrait (see Ref.\cite{DIFEQ}).  The
 equation can be re-written in the matrix form as
 
  \beq\label{SG1}
  \displaystyle{
\frac{d}{d\,z} \left( \begin{array}{c}
N\Lb z\Rb \\
N'_z\Lb z\Rb 
\end{array} \right)     \,\,=\,\,\left( \begin{array}{c}
N'_z\Lb z\Rb \\
\Lb 1 - \frac{1}{2\, a}\Rb\,N'_z\Lb z\Rb\,-\,\frac{1}{2\,a}\,N\Lb z\Rb\Big( 1\,\,-\,\,N\Lb z\Rb\Big)  
\end{array} \right)  }
\eeq 
 \eq{SG1} has two critical points:  $\Lb 0, 0\Rb$ and $\Lb 1, 0\Rb$. 
Near these critical points, \eq{SG1} can be re-written in the matrix form:
   \beq\label{SG11}
   \displaystyle{
\frac{d}{d\,z} \left( \begin{array}{c}
\Delta N\Lb z\Rb \\
\Delta N'_z\Lb z\Rb 
\end{array} \right)     \,\,=\,\,\mathbf{DF\Lb N_{\rm crit},N'_{\rm crit}\Rb } \left( \begin{array}{c}
\Delta N\Lb z\Rb \\
\Delta N'_z\Lb z\Rb 
\end{array} \right)  }
 \eeq 
 where $\Delta N $ denotes a  small deviation of $N$ in the vicinity of
  the critical point $N_{\rm crit}$. Matrices $\mathbf{DF}$ have the
 following forms:
      \beq\label{SG3}
      \displaystyle{
\mathbf{DF\Lb 0, 0 \Rb}\,\,=\,\, \left( \begin{array}{c c}
0 & 1 \\
-\frac{1}{2\,a} &  1 - \frac{1}{2\,a} 
\end{array} \right) ;~~~~~~~~~~~~~~~~\mathbf{DF\Lb 1, 0\Rb }\,\,=\,\, \left( \begin{array}{c c}
0 & 1 \\
\frac{1}{2\,a} &  1 - \frac{1}{2\,a} \end{array} \right) } \eeq 
      
\begin{figure}
\centering 
   \includegraphics[width=9cm]{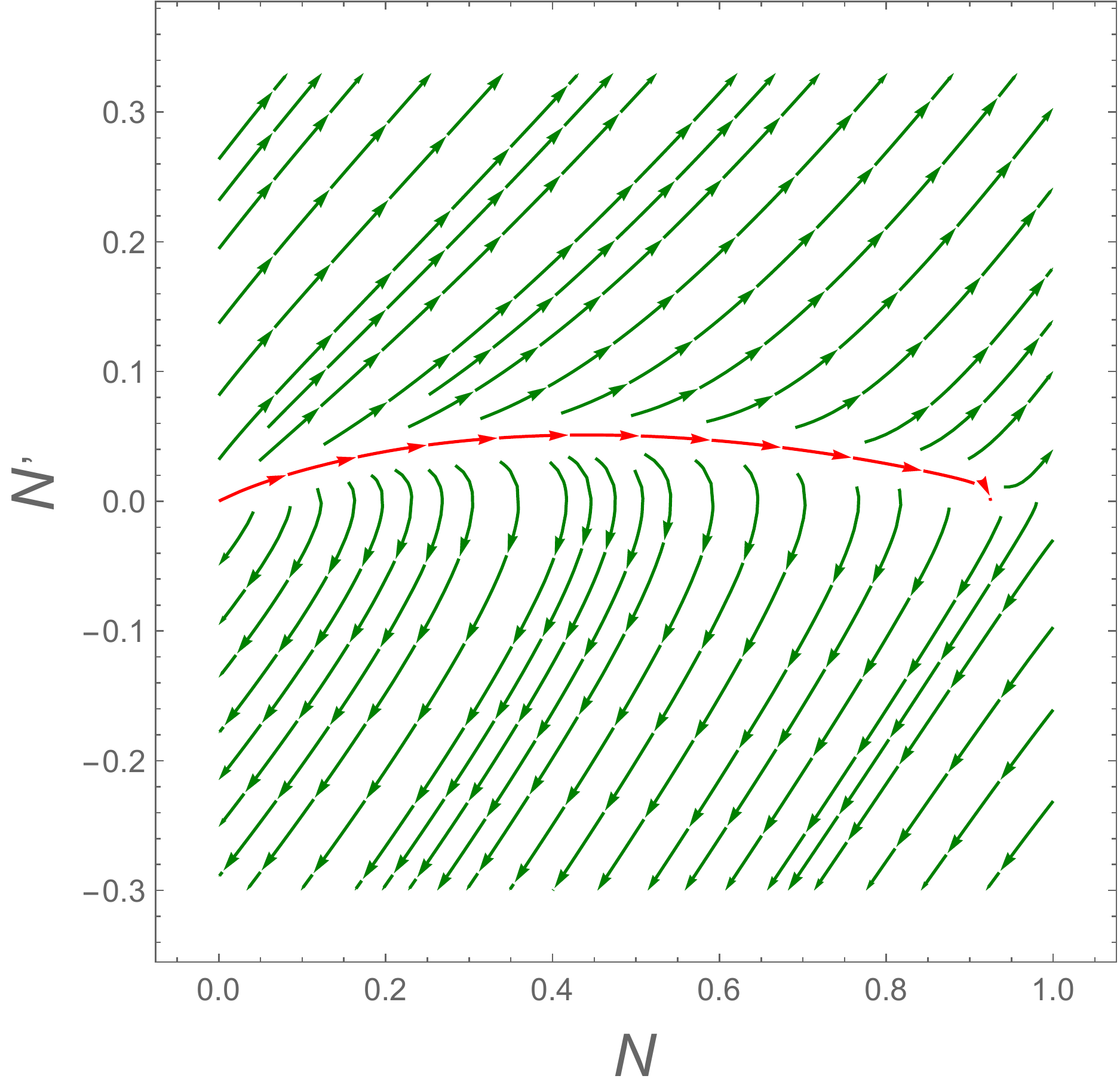}    
\caption{ The phase portrait of \eq{DLANL3} (or \eq{SG1}).The unique
 heteroclinic orbit for this equation is shown in red.}    
\label{str}
\end{figure}

   The eigenvalues of these matrices are equal to
 \bea \label{SG4}
 \mbox{critical point (0,0)} ~~&\to&~~ \Lb \lambda_1, \lambda_2\Rb \,\,=\,\,\Lb \sqrt{2}\,-\,1,\,\sqrt{2} \,-\,1
 \Rb;\nn\\
 \mbox{critical point (1,0)} ~~&\to&~~ \Lb \lambda_1, \lambda_2\Rb \,\,=\,\,\Lb1 , \, -3\,+\,2\,\sqrt{2}\Rb;     
 \eea        
 Therefore, the point (0,0) is the source from which our system
 originates, while the point (1,0) is a saddle point, since 
 only one of the eigenvalues is negative.  Hence, there exist
 one solution of our equation which gives
 
 \beq \label{SG5}
 \lim_{z \,\to\,-\,\infty}N\, \,=\,\,0; \lim_{z \,\to\,+\,\infty}\,N\, \,=\,\,1;  
 \eeq
 The path for this solution in the plane  $\Lb N(z),N_z(z)\Rb$ is shown
 in \fig{str} by a red line.
 
   Using \eq{SG4}, we can find solutions in the vicinity of the
 critical points. Indeed,  at $z \,\to \,- \infty$ the eigenfunctions
  of \eq{SG11}  in the vicinity of the critical point (0,0)  have the form:
   \beq \label{SG6}
  \displaystyle{
\mathbf{F_1}\,\,=\,\,\left( \begin{array}{c}
\sqrt{2}\,+\,1 \\
1 
\end{array} \right) \,e^{\lambda_1\,z};  ~~~~~~~~~~~~~  
\mathbf{F_2}\,\,=\,\,\left( \begin{array}{c}
1 \\
\sqrt{2}\,-\,1 
\end{array} \right) \,e^{\lambda_2\,z}; 
 }
\eeq    
 Therefore, the general solution at $z\,\to\,- \infty$ has the  form
    \beq \label{SG7}   
     \displaystyle{
\left( \begin{array}{c}
N\Lb z\Rb \\
N'_z\Lb z\Rb 
\end{array} \right)   \,\,=\,\,C_1\,\,\mathbf{F_1} \,\,\,+\,\,C_2\,\,\mathbf{F_2}  }
\eeq
Note that $\lambda_1=\lambda_2$, 
similarly, at $z\,\to\,\infty$ we have in the vicinity of the critical
 point (1,0):  \beq \label{SG8}
  \displaystyle{
\mathbf{F_1}\,\,=\,\,\left( \begin{array}{c}
1 \\
1 
\end{array} \right) \,e^{\lambda_1\,z};  ~~~~~~~~~~~~~  
\mathbf{F_2}\,\,=\,\,\left( \begin{array}{c}
1 \\
3\,-\,2\sqrt{2}
\end{array} \right) \,e^{\lambda_2\,z}; 
 }
\eeq 

and
  \beq \label{SG9}   
     \displaystyle{
\left( \begin{array}{c}
\Delta N\Lb z\Rb \\
\Delta N'_z\Lb z\Rb 
\end{array} \right)   \,\,=\,\,C_1\,\,\mathbf{F_1} \,\,\,+\,\,C_2\,\,\mathbf{F_2}  }
\eeq
Since  in vicinity of the critical point (1,0) $\lambda_1$   is positive, 
 we need to put $C_1=0$ and therefore, we have
 \beq \label{SG10}
           N\Lb z\Rb\,\,=\,\,1\,\,-\,C_2\exp\Lb -\, \Lb 3 \,-2\,\sqrt{2}\Rb\,z\Rb
           \eeq
    \paragraph{ Analytical solutions in the bounded regions:}     
       The previous analysis can be improved in two different
 kinematic regions: at small $N \,<\,1$, when non-linear corrections
 are small; and  at $N \,\to\,1$,  where we can develop the semi-classical
 approach.

   \eq{DLANL3} can be solved, by looking for the solution $ 
 \frac{d}{d \,z}\,N\Lb z\Rb\,\,=\,\,F\Lb N\Rb$. The equation takes the form:
     \beq \label{DLANL4}
 \h\,  \alpha\,\beta\,  \frac{d}{d N}\, F^2\Lb N\Rb\,\,+\,\,\Lb \alpha \,\,+\,\,\h\,\bas\,\beta\Rb\,F\Lb N\Rb =\,\, \h\,\bas \,\Big\{ N\,\,-\,\, N^2 \Big\}
     \eeq     
     
     For $N\,\ll\,1$ the solution of \eq{DLANL4}
 takes the form:
     \beq \label{DLANL5}
     F\Lb N\Rb\,\,=\,\,C_1\,N;~~~~  \alpha\,\beta\,C^2_1\,\,+\,\, \Lb \alpha \,\,+\,\,\h\,\bas\,\beta\Rb\,C_1\,\,=\,\,\h \,\bas
     \eeq
         
   As we have seen (see \eq{DLALONL60} and   \eq{DLANL2}), the natural
 choice of parameters $\alpha$ and $\beta$ is,  $\alpha\,=\,\,\h 
\,\bas\,\Lb 3 + 2\sqrt{2}\Rb\,\equiv\,a\,\bas$ and, $\beta \,=\,-1$,
 which leads to $C_1\,\, = \,\,\,\Lb \sqrt{2}\,-\,1\Rb$. The dependence
 of $N$ on $z$  can be  recovered from the equation:
   \beq \label{DLANL6}
   \frac{ d}{d \,z} N\,=\, F\Lb N\Rb \,=\, \Lb \sqrt{2}\,-\,1\Rb\,N 
   \eeq
which has the solution:
    \beq \label{DLANL7}
   \ln N\,\,=\,\,\Lb \sqrt{2}\,-\,1\Rb\,z + C_2;~~~~~~~~~N\Lb z\Rb\,\,=\,\,\Lb N_0\,=\,e^{C_2}\Rb \,e^ {C_1\,z}\,\,\,=\,\,\,N_0\,\Lb r^2\,Q^2_s\Lb Y\Rb\Rb^{\bar{\gamma}}
   \eeq
   with $Q^2_s\Lb Y\Rb\,\,=\,\,Q^2_s\Lb Y\,=\,0\Rb\exp\Lb \lambda\,Y\Rb$
 , where
   \beq \label{LAMBDA}
   \lambda\,\,=\,\,\h\bas \,\,\frac{1\,+\,\sqrt{2}}{\sqrt{2}\,-\,1 \,+\,\h \bas \Lb 1 \,+\,\sqrt{2}\Rb} ~~~~\mbox{and}~~~~~~~
   \bar{\gamma} \,\,\,=\,\,\,\sqrt{2} - 1 \,+\,\h\bas \Lb 1 \,+\,\sqrt{2}\Rb  
   \eeq
   
   Note that \eq{DLANL7} gives $N\Lb z\Rb$, as a
 function of $z\,=\,\alpha \,\eta \,\,+\,\,\xi$ (see \eq{DLANL3}) in
 the vicinity of the saturation scale.  As we have discussed, the energy
 variable $\eta$ leads to the \eq{DLANL1},  and \eq{EQQS} determines
  $Q^2_s(\eta)\,=\,Q^2_s\Lb \,0\Rb\exp\,\lambda_{\eta}\,\eta$. Therefore,
 to obtain $N\,=\,N_0\,\Lb r^2\,Q^2_s\Lb Y\Rb\Rb^{\bar{\gamma}}$ we need
 to re-calculate the variable $z$ in terms of $Y$ and  $\xi$: $z \,\,=
\,\,\lambda_{\eta}\,(Y\,+\,\xi)\,\,+\,\,\xi$. Hence, in \eq{DLANL7}
   \beq \label{LAMBDAY}
   C_1\,z\,\,\equiv\bar{\gamma}_{\eta}\,\,=\,\,C_1\Lb \lambda_{\eta}\,Y\,\,+\,\,(1 \,+\, \lambda_{\eta})\Rb\,\,=\,\,\Lb \lambda_Y\,Y   \,\,+\,\,\xi\Rb\,\bar{\gamma}
   \eeq
   with
     \beq \label{LAMBDAY1}
  \bar{\gamma}\,\,=\,\,C_1\,\Lb 1 \,+\,\lambda_\eta\Rb;~~~~~~\lambda_Y\,\,=\,\,\frac{\lambda_{\eta}}{1\,\,+\,\,\lambda_{\eta}}
     \eeq  
    \eq{LAMBDAY1} leads too the final expression in \eq{DLANL7} .

   It should be noted, that \eq{DLANL7} is the same as the solution of
 \eq{SG4} and \eq{SG6}, which we obtained in a different way.

\begin{figure}
\centering 
\begin{tabular}{c c c}
   \includegraphics[width=8cm]{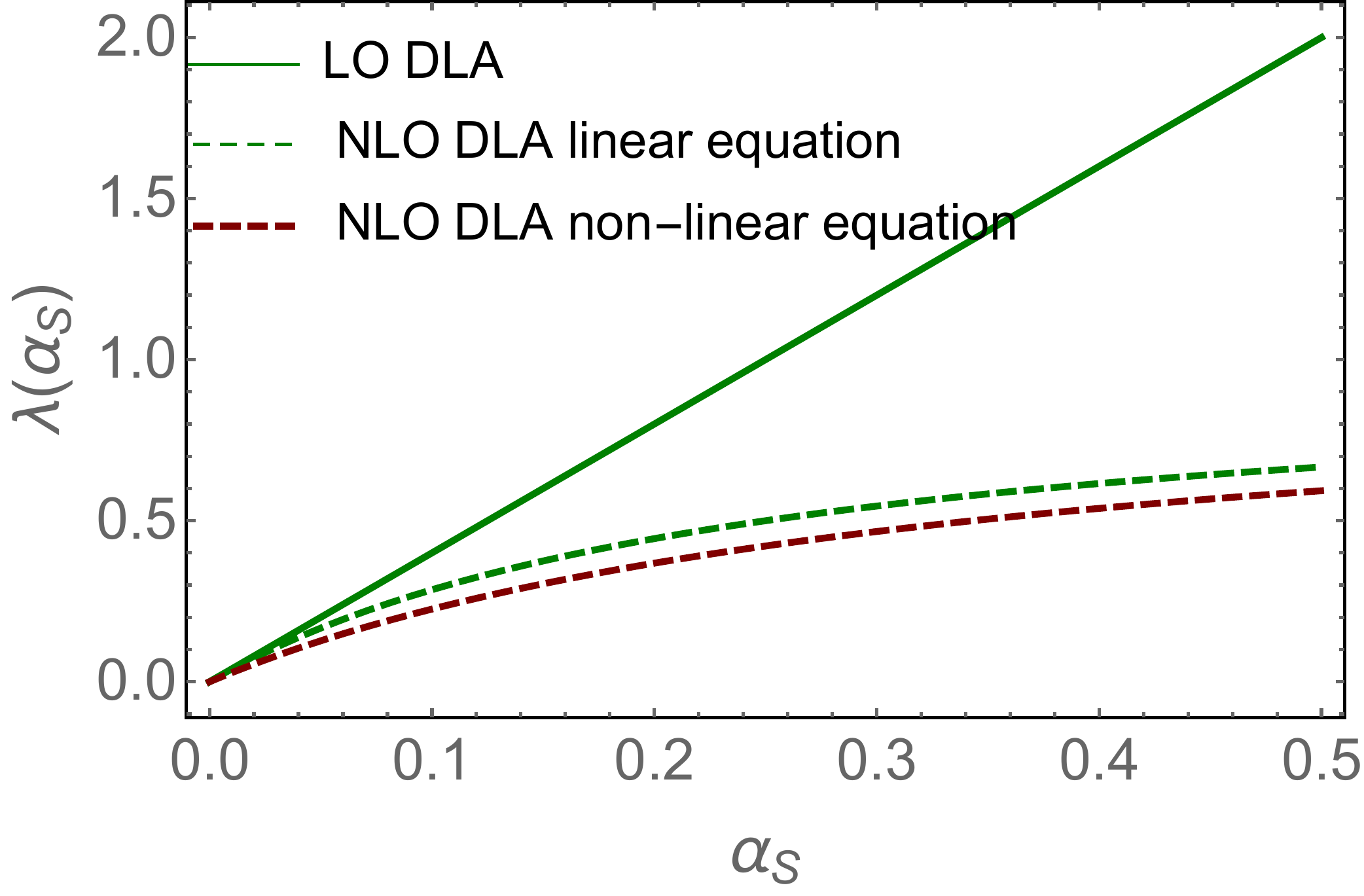} &~~~& \includegraphics[width=8cm]{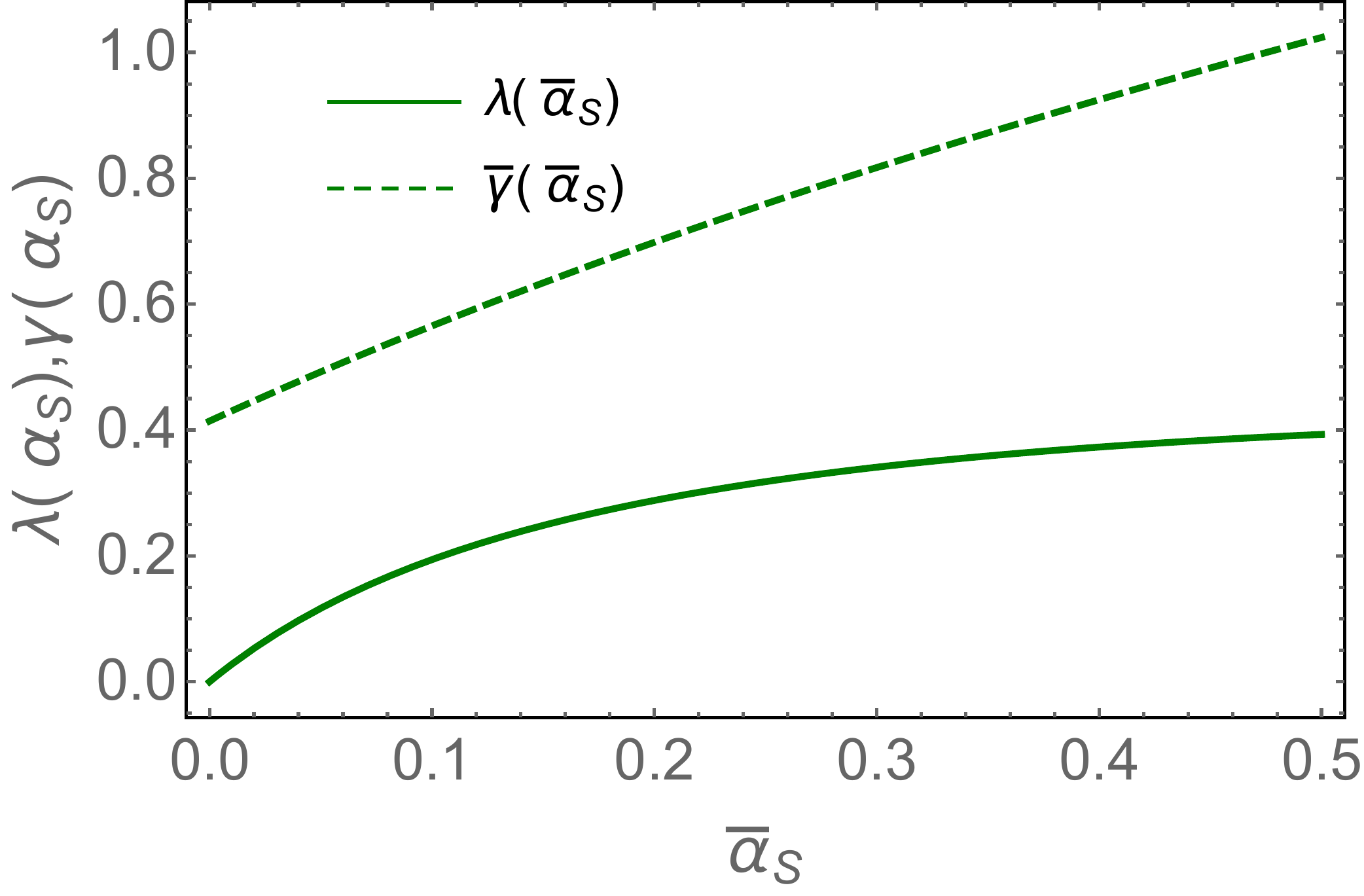}\\
   \fig{lam}-a &  & \fig{lam}-b\\
   \end{tabular}
\caption{ \fig{lam}-a: $ \lambda$ for Y dependence of the saturation
 scale: $Q^2_s\Lb Y\Rb\,\,=\,\,Q^2_s\Lb Y\,=\,0\Rb\exp\Lb \lambda\,Y\Rb$
 versus $\bas$ . The solid line corresponds to  $\lambda = 4 \bas$ in LO 
DLA,
 dashed green curve  describes the  NLO DLA (see \eq{PHASE})  and the
 dashed red curve is \eq{LAMBDA}. \fig{lam}-b: $\lambda(\bas)$ and
 $\bar{\gamma}(\bas)$ from \eq{ECDLA21}.}   
\label{lam}
\end{figure}

   In \fig{lam} we plot the dependence $\lambda $ on the values of $\bas$.
 One can see that \eq{LAMBDA} leads to $\lambda \,\approx\,0.3$ at rather
 large values of $\bas \approx 0.2$,  which is needed for the description
 of the HERA  experimental data

  In the kinematic region, where  $N \,\,\to\,\,1$ it is more convenient 
to solve
 \eq{DLANL3}  directly, by  looking for the solution in the form
 $N \,=\,1\,-\,\exp\Lb - \Omega(z)\Rb$, assuming that $\Omega\Lb z\Rb $
 is a smooth function of $z$ ($\Omega_{z z}\Lb z\Rb 
 \,\ll\,\Omega^2_{z}\Lb z\Rb  $). For such  a function
 $\Omega\Lb z \Rb$, \eq{DLANL3} can be re-written in the form:
  
  \beq \label{DLANLO8}
    -\, \alpha\,\beta\,  \Lb \frac{d}{d z} \,\Omega\Lb z\Rb\Rb^2\,\,+\,\,\Lb \alpha \,\,+\,\,\h\,\bas\,\beta\Rb\frac{d}{d \,z}\,\Omega\Lb z \Rb=\,\,\h\, \bas \,\Bigg\{ 1\,\,-\,\, \exp\Lb - \,\Omega\Lb z\Rb\Rb \Bigg\}
    \eeq
  
  Solving \eq{DLANLO8} we obtain
  
  \beq \label{DLANL9}
  \frac{d}{d \,z}\,\Omega\Lb z \Rb\,\,=\,\,\Lb \sqrt{2}\,-\,1\Rb\,\Lb -  1\,\pm\,\sqrt{2\,\,-\,\,\,\exp\Lb -   \Omega\Lb z \Rb\Rb}\Rb
  \eeq  
  
  Therefore, $\Omega\Lb z \Rb$  can be found from the equation
  \beq \label{DLANL10}
\frac{1}{ \sqrt{2}\,-\,1}\, \int^{\Omega\Lb z\Rb}_{\Omega_0}\frac{ d\,\Omega'}{-\,1\,\pm\,\sqrt{2 \,\,-\,\,\,\exp\Lb - \Omega'\Rb}}\,\,=\,\,z
  \eeq
  
  We choose the sign $+$ in \eq{DLANL10} since we are looking for 
 $\Omega(z)$ 
which is positive at large $z$, leading to $N \,\leq\,1$.
  
 Using \eq{DLANL10}  we can evaluate $\Omega'_z\Lb z\Rb$ and
 $\Omega''_{z z}\Lb z\Rb$:
 \bea \label{DLANL101} 
 \Omega'_z\Lb z\Rb\,\,&=&\,\, -\,\Lb \sqrt{2}\,-\,1\Rb\,\Lb 1\,\,-\,\,\,\sqrt{ 2\,-\,e^{- \Omega(z)}}\Rb\,;\nn
 \\
 \Omega''_{z z}\Lb z\Rb \,\,&=&\,\,\h\,\Lb \sqrt{2}\,-\,1\Rb^2\,\Bigg( 1\,\,-\,\,\frac{1}{\sqrt{ 2\,-\,e^{- \Omega(z)}}}\Bigg)\,e^{ -\,\Omega\Lb z\Rb}\,;
 \eea 
    
 The integral over $\Omega'$ can be taken then,  and \eq{DLANL10}has the 
form:
  \bea \label{DLANL102}
z\,&=&\,-\, \frac{1}{\sqrt{2}-1}\,\Bigg( -\,\Omega\,-\log \left(1-e^{-\,\Omega}\right)+2 \sqrt{2} \tanh ^{-1}\left(\sqrt{1-\frac{e^{-\,\Omega}}{2}}\right)-2 \tanh ^{-1}\left(\sqrt{2-e^{-\,\Omega}}\right)\nn\\
 &  & ~~~~~~~~~~~~~~~~\,\Omega_0\,+\log \left(1-e^{-\,\Omega_0}\right)\,-\,2 \sqrt{2} \tanh ^{-1}\left(\sqrt{1-\frac{e^{-\,\Omega_0}}{2}}\right)\,+\,2 \tanh ^{-1}\left(\sqrt{2-e^{-\,\Omega_0}}\right)
\Bigg)\eea  
  where $\Omega_0$ is the initial value of $\Omega $ at $z=0$.

  At large $z$,  $\,\Omega\Lb z \Rb\,\,\to\,\,\Lb \sqrt{2} - 1\Rb\,z $,
 while at small $z$,
  $\Omega\Lb z \Rb\,\,=\,\,\exp\Lb\h\Lb \sqrt{2} - 1\Rb\,z\Rb$. Note,
 that we obtain  similar behaviour of the amplitude at $z \,\to\,0$ 
 as in \eq{DLANL7}, but with a different power of $z$.
  
\begin{figure}
\centering
\begin{tabular}{c  c  c}   
   \includegraphics[width=6cm]{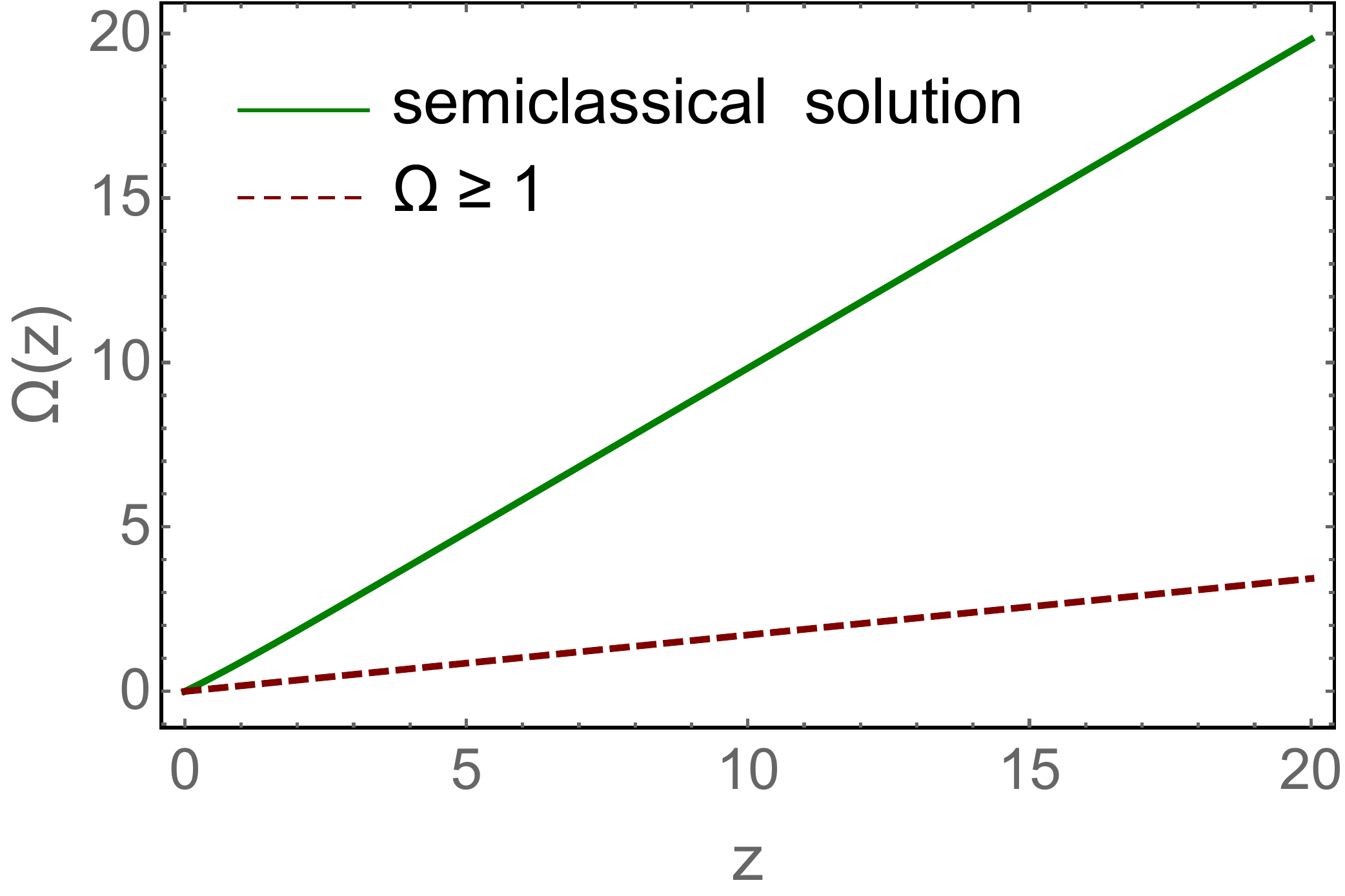}&   \includegraphics[width=6cm]{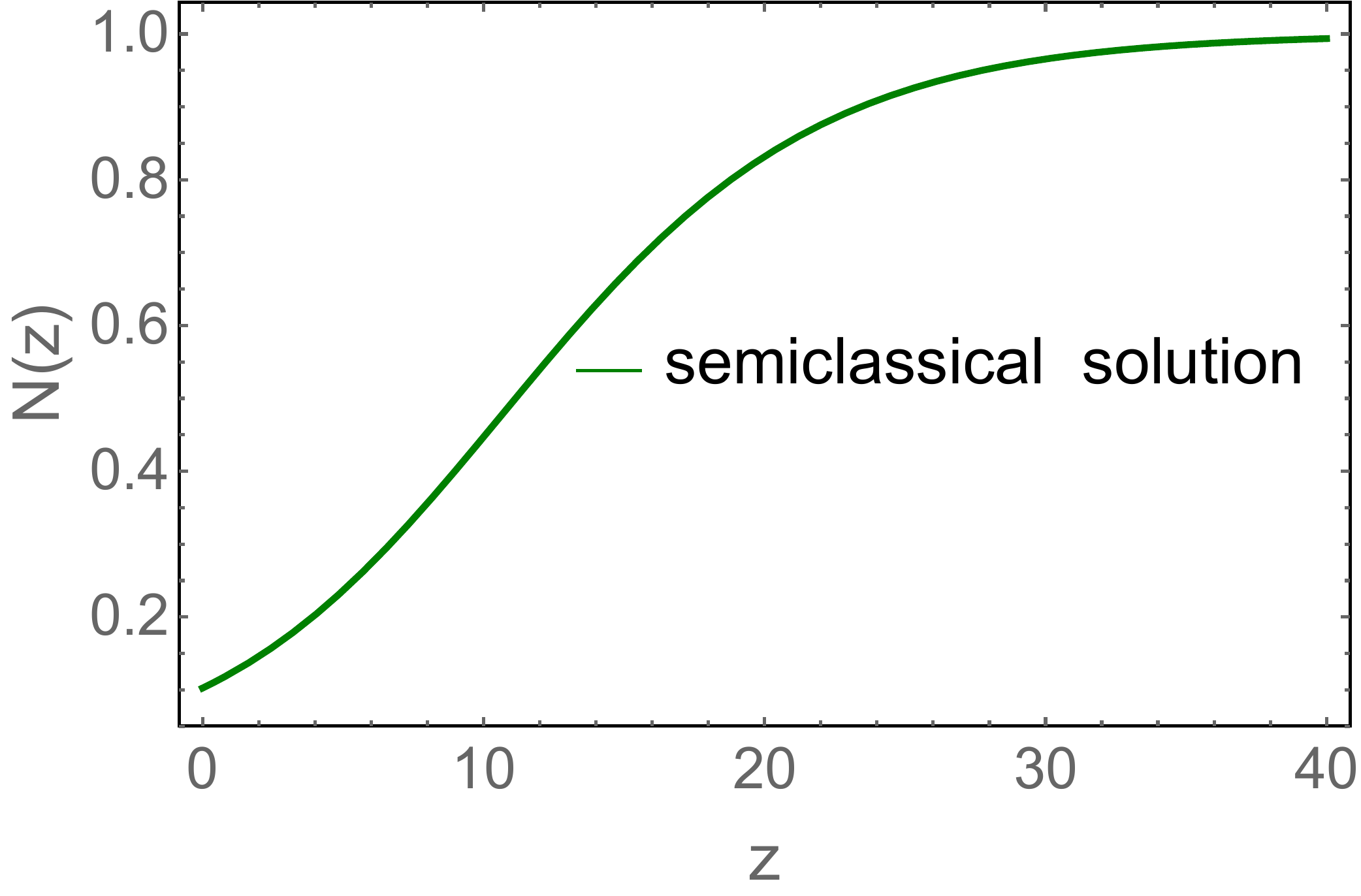}& \includegraphics[width=6.5cm]{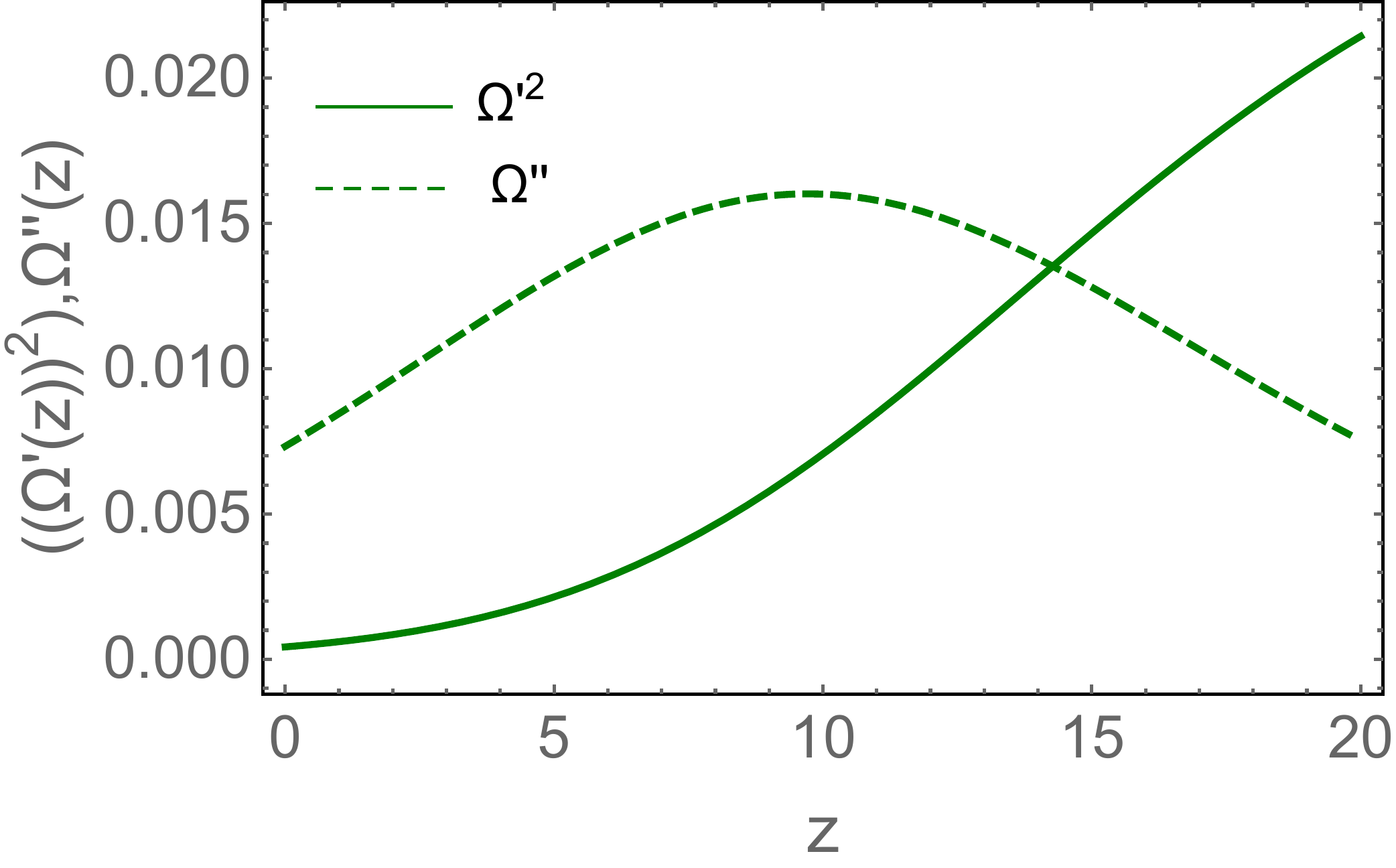}     \\
   \fig{solnum}-a& \fig{solnum}-b&\fig{solnum}-c\\
   \end{tabular}
    \caption{ \fig{solnum}-a: Function $\Omega\Lb z\Rb\,$ versus $z$:
 the solid line shows the solution of \eq{DLANL10} with the initial
 value of $\Omega_0 = 0.1$.
 The dotted line describes the solution at  $N \,\to\,1$ which is
 $\Omega\,=\,
 \frac{1}{8}\Lb \sqrt{17} - 3\Rb z$. The dotted line describes the
 solution at small $N \,<\,1$ (see \eq{DLANLO7}). \fig{solnum}-b: 
 Scattering amplitude $N \,=\,1 - \exp\Lb - \Omega\Rb$ versus $z$.
 \fig{solnum}-c: Functions $\Omega'_z\Lb z\Rb\,$ and 
 $\Omega''_{zz}\Lb z\Rb\,$ versus $z$.}    
\label{solnum}
\end{figure}

  In \fig{solnum} we plot the solution of \eq{DLANL11}. Note, that
 \fig{solnum}-c shows that $\Omega''_{zz}\,\ll\,\Omega'^2_{z}  $
 only for $z\,>\,12 \div 15$. Therefore, for $z \,<\,12 \div 15$
 we have to solve the following  equation:
     \beq \label{DLANL12}
    -\, \alpha\,\beta\,\Bigg\{ - \frac{d^2}{d \,z^2} \,\Omega\Lb z\Rb\,\,+\,\, \Lb \frac{d}{d z} \,\Omega\Lb z\Rb\Rb^2 \Bigg\}\,\,+\,\,\Lb \alpha \,\,+\,\,\h\,\bas\,\beta\Rb\frac{d}{d \,z}\,\Omega\Lb z \Rb=\,\,\h\, \bas \,\Bigg\{ 1\,\,-\,\, \exp\Lb - \,\Omega\Lb z\Rb\Rb \Bigg\}
    \eeq

   \paragraph{Practical way to obtain estimates for the scattering amplitude:}
     
\begin{figure}
\centering 
   \includegraphics[width=9cm]{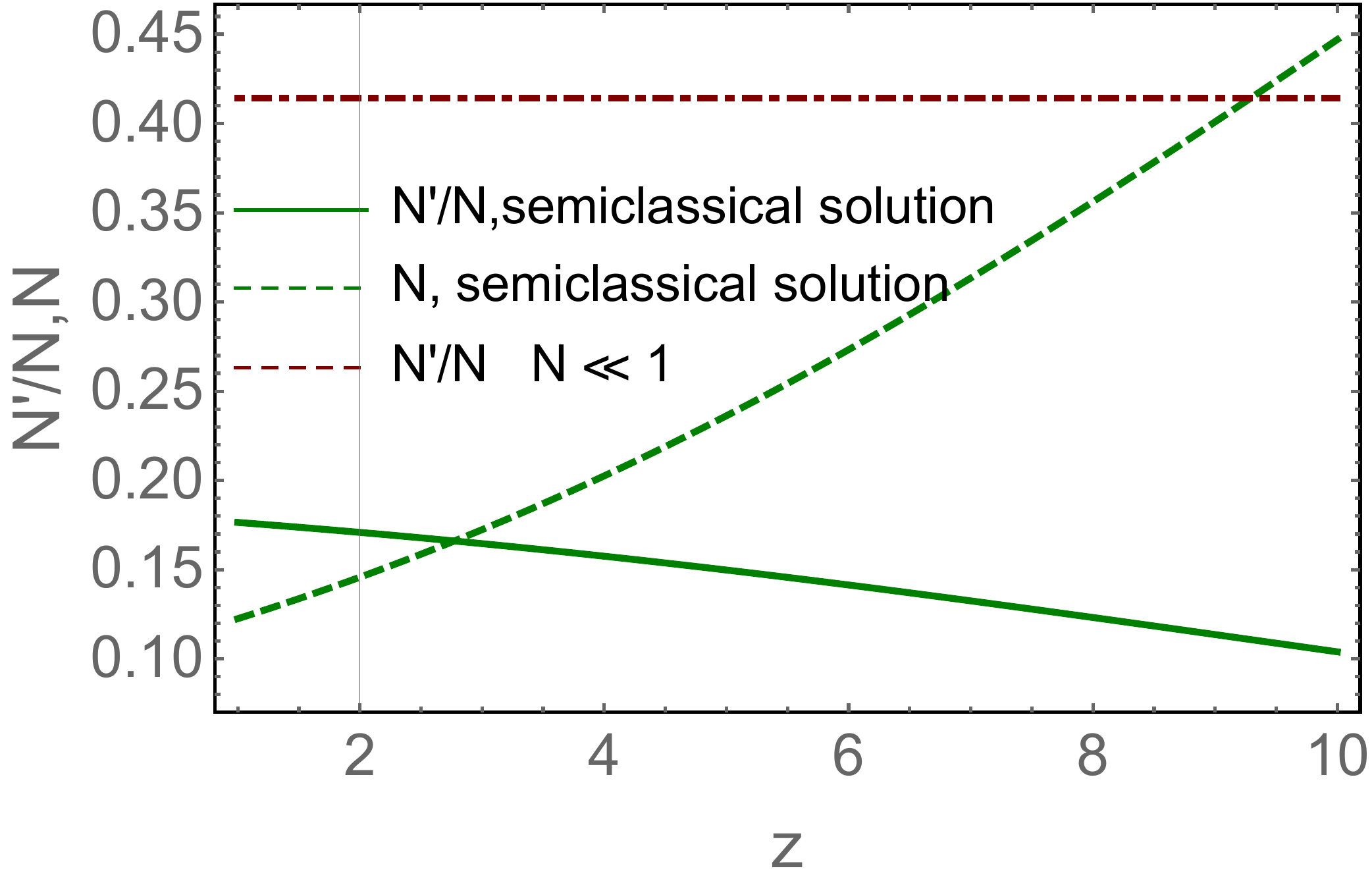}    
\caption{   $N'_z/N$ and $N$ for the semiclassical solution
 of \eq{DLANLO8} are shown in green,
while in red we show the values of $N'_z/N$ for the solution
 of \eq{DLANL7} for small $N$.}    
\label{difi}
\end{figure}

     
     The difficulties in solving the general equations (see \eq{DLANL3}
 and \eq{DLANL12}), are  that the critical point (1,0)  is not a stable
 point,  but only a saddle point. This means that there exists  only one
 path which gives the solution with $N\,\to\,1$ at large $z$ (see
 \fig{str}). It is  difficult to find this trajectory analytically
 or/and numerically. These difficulties are illustrate by \fig{difi} where 
  $N'_z/N$ and $N$ for the semiclassical solution of \eq{DLANLO8}  
 are shown in green, while the values of  $N'_z/N$   for small $N$,
  are plotted in red. One can see that it is  not possible to find a
 function which provides a smooth matching to these two solutions. The
 only possibility is to find the exact solution, that covers a large
 region of $z$ .

   The one way out,  comes from the hope that the solution at small $N$,
 is sufficient to cover the region of $\gamma$ from $\bar{\gamma} $ to
 $\gamma=\h$, where we have to use \eq{DLANL3}. We believe that
 for smaller $\gamma$ we need to use the different expression for
 the eigenvalues $\omega\Lb \gamma\Rb$, which is given by \eq{OMGA0}.

   \subsubsection{Energy conservation in DLA}

   
   We need to consider \eq{OMNLOKMRS}, which follows from  \eq{KMRSOM} in
 the region $\gamma\,\to\,1$,  for  taking   energy
 conservation into account.  Resolving this equation with respect to 
$\gamma$ we
 obtain for the amplitude $\tilde{N}\Lb \xi', Y \Rb$:
   \beq \label{ECDLA1}
1 \,-\,\gamma\,\,\,=\,\,\,\bas\Bigg( \frac{1}{\omega} \,\,-\,\,1\Bigg) \,\,-\,\,\omega
 \eeq   
         which leads to the following equation
  \beq \label{ECDLA2}
 \frac{\partial^2}{\partial \,\eta\,\partial\,\xi'}\tilde{N}\Lb \xi', \eta; b \Rb\,\,+\,\,\bas \frac{\partial}{\partial \,\eta}\tilde{N}\Lb \xi', \eta; b \Rb=\,\, \bas \, \tilde{N}\Lb \xi', \eta; b\Rb\,;\eeq  
 instead of \eq{DLANLO3}.
Repeating the same procedure as in section III-C-3,  the following
  non-linear equation can be written:   
\beq \label{ECDLA20}
 \frac{\partial^2}{\partial\,\eta\,\partial\,\xi'} N\Lb \xi', \eta \Rb \, \,+\,\,\Lb 1\,+\,\bas\Rb  \frac{\partial}{\partial\,\eta} N\Lb \xi', \eta \Rb 
  \,\,+\,\,\h\,\bas\,\frac{\partial}{\partial\,\xi'} N\Lb \xi', \eta\Rb =\,\, \h\,\bas \,\Bigg\{ N\Lb \xi', \eta\Rb\,\,-\,\, N^2\Lb \xi', \eta\Rb \Bigg\}.
     \eeq 

 Neglecting the $N^2$ term in \eq{ECDLA20},  we have a linear equation 
which determines the value of the saturation scale 
 $Q^2_s\Lb \eta\Rb\,\,=\,Q^2_s\Lb \,0\Rb\exp\,\lambda_{\eta}\,\eta$
  and $\bar{\gamma}_{\eta}$:
\beq \label{QSETA}
\bar{\gamma}_{\eta}\,\,=\,\,\sqrt{ 2\,+\,\bas}\,
\,-\,\,1;~~~~\lambda_{\eta}\,\,=\,\,\h\,\bas \frac{
 \sqrt{2\,+\,\bas}}{3 \sqrt{ 2\,+\,\bas}\,-\,4 \,
+\,2\,\bas\Lb  \sqrt{2 \,+\,\bas}\,-\,1\Rb}
\eeq

  Tedious but simple calculations following the procedure of
 section III-C lead to

\bea \label{ECDLA21}
\ln\Lb Q^2_s\Lb Y\Rb/ Q^2_s\Lb Y=0\Rb\Rb\,\,&=&\,\,\lambda\,Y~~~ \mbox{with} \,\,\,\,\lambda\,\,=\,\,\frac{\bas\,\Lb \sqrt{2\, +\, \bas} \,+\,1\Rb}{\sqrt{1\,+\,\bas} \Lb 2 \,+\,3\bas \,-\,\sqrt{2\, +\,\bas}\Rb };\nn\\
\bar{\gamma}\,\,&=&\,\sqrt{2\,+\,\bas} \,-\,1 \,\,+\,\,\bas \frac{1\,+\,\sqrt{2\,+\,\bas}}{2\,\Lb 1\,+\,\bas\Rb}\eea

The values of $\lambda(\bas)$ and $\bar{\gamma}(\bas)$ are plotted 
in \fig{lam}-b.

Introducing a new variable $z\,\,=\,\,\alpha\,\eta\,\,+\,\,\beta\,\xi'$, 
 this equation has the form:

\beq \label{ECDLA22}
  \alpha\,\beta\,  \frac{d^2}{d\,z^2} N\Lb z \Rb \, \,+\,\,\Lb \alpha\,\Lb 1\,+\,\bas\Rb  \,\,+\,\,\h\,\bas\Rb\,\frac{d}{d\,z} N\Lb z \Rb =\,\, \h\,\bas \,\Bigg\{ N\Lb z\Rb\,\,-\,\, N^2\Lb z\Rb \Bigg\}
     \eeq 
 A natural choice of $\alpha $ and $\beta$ is $\alpha\,\,=\,
\,\lambda_{\eta}(\bas)$ and $\beta\,=\,-1$. Considering
$N'_z\Lb z \Rb\,=\,F\Lb N\Rb$ we can re-write \eq{ECDLA22} in the form
\beq \label{ECDLA3}
  2 \alpha\,\beta\,  \frac{d}{d N}\, F^2\Lb N\Rb\,\,+\,\,\Lb \alpha\,\Lb 1\,+\,\bas\Rb  \,\,+\,\,\h\beta\bas\Rb\,F\Lb N\Rb =\,\, \h\,\bas \,\Bigg\{ N \,\,-\,\, N^2  \Bigg\}
     \eeq

    For $N\,\ll\,1$,  the solution of \eq{ECDLA3} can be obtained in
 the same way as the solution of \eq{DLANL5}:
       \beq \label{ECDLA4}
     F\Lb N\Rb\,\,=\,\,C_1\,N;~~~~  \alpha\,\beta\,C^2_1\,\,+\,\,\, \Lb \alpha\Lb 1 + \bas\Rb \,\,+\,\,\h\,\frac{\bas\,\beta}{1+\bas}\Rb\,C_1\,\,=\,\,\h \bas
     \eeq
      
    The solution to \eq{ECDLA4} can be written as follows:
    \beq \label{ECDLA5}
N\Lb z\Rb\,\,=\,\,\,N_0\,\Lb r^2\,Q^2_s\Lb Y\Rb\Rb^{\bar{\gamma}}
   \eeq
   with  $\lambda $ and $\bar{\gamma}$  given in \eq{ECDLA21}.  
  \section{NLO BFKL kernel in the saturation domain}
  \begin{boldmath}
  \subsection{The kernel in $\gamma$-representation}
  \end{boldmath}
  
The BFKL kernel of \eq{CHI} includes the summation over all twist
 contributions. In the simplified approach we restrict ourselves 
 to the leading twist term only, which has the form\cite{LETU} 
\bea \label{SIMKER}
\chi\Lb \gamma\Rb\,\,=\,\, \left\{\begin{array}{l}\,\,\,\frac{1}{\gamma}\,\,\,\,\,\,\,\,\,\,\mbox{for}\,\,\,\tau\,=\,r Q_s\,>\,1\,\,\,\,\,\,\mbox{summing} \Lb \ln\Lb r Q_s\Rb\Rb^n;\\ \\
\,\,\,\frac{1}{1 \,-\,\gamma}\,\,\,\,\,\mbox{for}\,\,\,\tau\,=\,r Q_s\,<\,1\,\,\,\,\,\mbox{summing}
\Lb \ln\Lb1/(r\,\Lambda_{\rm QCD})\Rb\Rb^n;\\  \end{array}
\right.
\eea
instead of the full expression of \eq{CHI}.

In the previous section we specified how we changed the kernel in the
 perturbative QCD region, taking into account the NLO corrections.  We
   now desire to find   how we need to change the kernel $1/\gamma$
  for $\tau\,=\,r Q_s\,>\,1$ (in the saturation region).  For this
 purpose we re-write \eq{KMRSOM}  in the vicinity of $\gamma \to 0$,
  where it has the form:
 \beq \label{OMGA0}
 \omega_{<}\,\,=\,\,\bas \Lb 1 \,-\,\omega_{<}\Rb \Bigg\{ \frac{1}{\gamma} \,\,-\,\,\omega_{<}\Bigg\}
 \eeq
  This eigenvalue describes the behaviour of  $\omega^{\rm KMRS}$ for
 $\gamma \leq \h$ (see \fig{omi}).   We start from the simplified expression
 for \eq{OMGA0}:
   \beq \label{OMGA01}
 \omega_{<}\,\,=\,\,\bas \Bigg\{ \frac{1}{\gamma} \,\,-\,\,\omega_{<}\Bigg\}
 \eeq  
  Resolving \eq{OMGA01} with respect of $\gamma$ we obtain
  
  \beq \label{KERGA1}
  \gamma\,\,=\,\,\frac{\bas}{1 + \bas }\, \frac{1}{\omega}
  \eeq
  which differs from the behaviour of the LO kernel, only by  replacing 
 $\bas \,\to\,\bas/\Lb 1 \,+\,\bas\Rb$. Therefore, we can discuss the
 non-linear equation in the LO,   substituting $\bas/\Lb 1 \,+\,\bas\Rb$
 in place of $\bas$, at the final stage.  
    
  \subsection{The non-linear equation}
   In the saturation region 
 where $\tau\,\,>\,\,1$,  the  logarithms
   originate from the decay of a large size dipole, into one small
 size dipole  and one large size dipole\cite{LETU}.  However, the size of the
 small dipole is still larger than $1/Q_s$. This observation can be
 translated in the following form of the kernel in the LO
\bea \label{K2}
\frac{\bas}{2 \pi}\int \, \displaystyle{K\Lb \vec{x}_{01};\vec{x}_{02},\vec{x}_{12}\Rb}\,d^2 x_{02} \,&\rightarrow&
\,\frac{\bas}{2}\, \int^{x^2_{01}}_{1/Q^2_s(Y,b)} \frac{ d x^2_{02}}{x_{02}^2}\,\,+\,\,
\frac{\bas}{2}\, \int^{x^2_{01}}_{1/Q^2_s(Y, b)} \frac{ d |\vec{x}_{01}  -
 \vec{x}_{02}|^2}{|\vec{x}_{01}  - \vec{x}_{02}|^2}\,\,\nn\\
 &=&\,\,\frac{\bas}{2}\, \int^{\xi}_{-\xi_s} d \xi_{02} \,\,+\,\,\frac{\bas}{2}\, \int^{\xi}_{-\xi_s} d \xi_{12}\eea
 where $\xi_{ik} \,=\,\ln\Lb x^2_{ik} Q_s^2(Y=Y_0)\Rb$
 and $\xi_s\,=\,\ln\Lb Q^2_s(Y)/ Q_s^2(Y=Y_0)\Rb$.

Inside the saturation region the BK equation of the LO  takes the form
\beq \label{BK2}
\frac{\partial^2 \widehat{N}\Lb Y, \x; \vec{b}\Rb}
{ \partial Y\,\partial \xi}\,\,=\,\, \bas \,\left\{ \Lb 1 
\,\,-\,\frac{\partial \widehat{N}\Lb Y, \xi; \vec{b}
 \Rb}{\partial  \xi}\Rb \, \widehat{N}\Lb Y, \xi;
 \vec{b}\Rb\right\}
\eeq
where 
 $\widehat{N}\Lb Y, \xi; \vec{b}\Rb\,\,=\,\,\int^{\xi} d \xi'\,N\Lb Y,
 \xi'; \vec{b}\Rb$ .
    
  For the NLO kernel of \eq{KERGA1},  \,  \eq{BK2} takes the form:
  \beq \label{BK3}
\frac{\partial^2 \widehat{N}\Lb Y, \xi; \vec{b}\Rb}
{ \partial Y\,\partial \xi}\,\,=\,\, \frac{\bas}{1\,+\,\bas} \,\left\{ \Lb 1 
\,\,-\,\frac{\partial \widehat{N}\Lb Y, \xi; \vec{b}
 \Rb}{\partial  \xi}\Rb \, \widehat{N}\Lb Y, \xi;
 \vec{b}\Rb\right\}
\eeq  
  
\subsection{The solution}

For solving this equation we introduce function $\Omega\Lb Y;
 \xi, \vec{b}\Rb$\cite{LETU}
\beq \label{SOL1}
N\Lb Y, \xi \Rb\,\,=\,\,1\,\,-\,\,\exp\Lb - \Omega\Lb Y, \xi\Rb\Rb
\eeq
Substituting \eq{SOL1} into \eq{BK3} we  reduce it to the form

\begin{subequations}
\bea \label{SOL2}
&&\frac{ \partial \Omega\Lb Y, \xi\Rb}{ \partial Y} \,\,=\,\,\frac{\bas }{1 + \bas}\widetilde{N}\Lb Y, \xi\Rb;~~~\frac{ \partial^2 \Omega\Lb Y, \xi\Rb}{ \partial Y\,\partial \xi}  \,\,\,=\,\,\frac{\bas }{1 + \bas}\Bigg( 1 -\,\exp\Lb - \Omega\Lb Y, \xi\Rb\Rb\Bigg);\label{SOL02}\\
&&~\frac{ \partial^2 \Omega\Lb \xi_s; \zeta\Rb}{ \partial \xi_s\,\partial \xi}  \,\,\,=\,\,\frac{\bas}{\lambda(\bas)\,\Lb 1\,\,+\,\,\bas\Rb}\Bigg( 1 -\,\exp\Lb - \Omega\Lb \xi_s; \zeta\Rb\Rb\Bigg)
\,\,\equiv\,\,\sigma\Bigg( 1 -\,\exp\Lb - \Omega\Lb \xi_s; \zeta\Rb\Rb\Bigg)\label{SOL2}\eea
\end{subequations}
where $\lambda$  is given by   \eq{LAMBDA}. The variable $\xi_s$ is
 defined as 
\beq \label{XIS}
\xi_s\,\,=\,\,\ln\Lb Q^2_s\Lb Y\Rb/Q^2_s\Lb Y=0; \vec{b},\vec{R}\Rb\Rb\,\,=\,\,\lambda \,Y
\eeq
The use of this variable indicates the main idea of our approach: we
 wish to match the solution of the non-linear \eq{SOL2} with the solution
 of the non-linear \eq{ECDLA3}. However, we assume that for $\bar{\gamma}
 \,\geq\,\h$ we need the solution for \eq{ECDLA3}  for $N\,\,<\,\,1$ ,
 where it has the form $N \,\,=\,\,N_0 \,\exp\Lb \bar{\gamma} \,z\Rb$
 with $z$
\beq \label{Z}
z\,\,=\,\,\xi_s\,\,+\,\,\xi
\eeq
and $\bar{\gamma}$ is determined by \eq{LAMBDA}.

\eq{SOL2} has a traveling wave solution (see formula {\bf 3.4.1.1} of
  Ref.\cite{MATH}). For \eq{SOL2} in the canonical form:
\beq \label{SOL3}
\frac{ \partial^2 \Omega\Lb \xi_s; \tilde{ \zeta}\Rb}{ \partial t^2_+} \,\,-\,\,\frac{ \partial^2 \Omega\Lb \xi_s; \tilde{\zeta}\Rb}{ \partial t^2_-} \,\,\,=\,\,\sigma\,\,\Bigg( 1 -\,\exp\Lb - \Omega\Lb \xi_s; \tilde{\zeta} \Rb\Rb\Bigg),
\eeq
with $t_{\pm} = \xi_s \pm  \xi $, the solution takes the form:
\beq \label{SOL4}
\int^{\Omega}_{\Omega_0}\frac{d \Omega'}{\sqrt{ C_1 + \frac{2}{(\mu^2 - \kappa^2)} \sigma \Lb \Omega' + \exp\Lb-\Omega'\Rb\Rb}}\,\,=\,\,\mu t_+ + \kappa t_-  + C_2
\eeq
where all constants have to be determined  from the initial and boundary 
conditions of \eq{IC}. First we see that $C_2=0 $ and $\kappa =0$.
 From the condition $\Omega'_z/\Omega \,\,=\,\,\bar{\gamma}$ at $t_+ =
 0$ we can find $C_1$. Indeed, differentiating \eq{SOL4} with respect to 
$t_+$ one
 can see that at $t_+= 0$ we have:
\beq \label{SOLSET}
\frac{d \Omega}{d t_+}|_{t_+ = 0}\,\frac{1}{\sqrt{ C_1 \,\,+\,\,\frac{2 \,\sigma}{\mu^2}\Lb 1\,\,+\,\,\h\,\Omega^2_0\Rb}}\,\,=\,\,\mu
\eeq

From     \eq{SOLSET} one can see that choosing 

\beq \label{SOLSET1}
C_1\,\,=\,\,- \,\frac{2 \sigma}{\mu^2} \,\,+\,\,\Omega^2_0 \Lb 1\,\,-\,\,\frac{ \sigma}{\mu^2}\Rb;~~~~~\mbox{and}\,~~~\mu \,\,=\,\,\bar{\gamma} 
\eeq

we satisfy the initial condition $\frac{d \ln\Lb \Omega\Rb}{d z}|_{t_+ = 0}
 =\,\bar{\gamma}$ of \eq{LAMBDA}.

Finally, the solution of \eq{SOL4} can be re-written in the following
 form for $\Omega_0 \ll 1$:
\beq \label{SOL5}
\int^{\Omega}_{\Omega_0}\frac{d \Omega'}{\sqrt{ \Omega^2_0\,\Lb1\,\,-\,\,\frac{\sigma}{\bar{\gamma}^2}\Rb\,\,+\,\,\frac{2\,\sigma}{\bar{\gamma}^2}\Lb -1\,\,+\,\,\Omega'\,\,+\,\,\,e^{ - \,\Omega'}\Rb
}}\,\,=\,\, \bar{\gamma}\,z \eeq
For $\Omega \to \Omega_0$ and if $\Omega_0 \ll 1$, \eq{SOL5} can be 
 solved explicitly giving
\beq \label{SOL7}
\Omega\,\,=\,\,\Omega _0	\Bigg\{ \cosh \left(\sqrt{\sigma } \Lb\xi_s + \xi \Rb \right)\,\,+\,\,\frac{\bar{\gamma}}{\sqrt{\sigma}}\, \sinh \left(\sqrt{\sigma}\Lb  \xi_s + \xi\Rb\right)\Bigg\}\eeq
 \eq{SOL7} gives the solution which depends only on one variable
 $z\,\,=\,\,\xi_s + \xi$,
and satisfies the initial conditions of \eq{IC}.

At large $z$ we obtain the solution\cite{LETU}:
\beq \label{SOL8}
\Omega\Lb z\Rb\,\,=\,\,\frac{\sigma}{2}\,z^2\,\,+\,\,{\rm Const}
\eeq
or in terms of the amplitude
\beq \label{SOL9}
N\Lb z\Rb\,\,=\,\,1\,\,-\,\,{\rm Const}\,e^{-\,\frac{\sigma}{2}\,z^2}
\eeq
  We wish  to stress that  \eq{SOL9} reproduces the 
 asymptotic solution to \eq{NLOBK}, which has been derived in
 Refs.\cite{CLMP,XCWZ}, for fixed $\bas$.

It should be noted that both solutions of \eq{SOL7} and \eq{SOL8}
 can be derived directly from \eq{SOL3} assuming $1\,-\,\exp\Lb -
 \Omega\Rb\,\,\to\,\Omega$ and $1\,-\,\exp\Lb - \Omega\Rb\,\,\to\,1$ for
 small  $z$ and large $z$, respectively.

\subsection{Non-linear equation:  NLO + energy conservation}

 In this section we discuss the general form of the eigenvalues
 which is given by \eq{OMGA0}. We obtain the following solution for
 $\omega$ from this  equation.
 \beq \label{ECN1}
\omega\Lb \gamma\Rb\,\,=\,\,\widetilde{\as}\frac{1}{\gamma \,+\,\widetilde{\as}}\,\,=\,\,\widetilde{\as}
\int^\infty_0 d \,\rho\,\,e^{ - \Lb \gamma\,+\,\widetilde{\as}\Rb\,\rho}
\eeq 
where $\widetilde{\as} \,=\,\bas/(1 + \bas)$ and $\rho\,=\,\xi\,-\xi'\,
=\ln\Lb x^2_{01} \,Q^2_s(Y)\Rb\,\,-\,\,\ln\Lb x^2_{02} \,Q^2_s(Y)\Rb$
 (see \eq{DLANLO4}).

Therefore, the general solution for linear equation takes the form:

\beq \label{ECN2}
N\Lb Y, \xi\Rb\,\,=\,\,\int^{\epsilon + i\infty}_{\epsilon -  i\infty} \frac{ d \gamma}{2 \pi\,i}e^{ \widetilde{\as}\frac{1}{\gamma \,+\,\widetilde{\as}}\,Y\,\,+\,\,\gamma\,\xi}\, n_{\rm in}\Lb \gamma\Rb\,\,\equiv\,\,\int^{\epsilon + i\infty}_{\epsilon -  i\infty} \frac{ d \gamma"}{2 \pi\,i}e^{ \widetilde{\as}\frac{1}{\gamma" }\,Y\,\,+\,\,\gamma"\,\xi\,\,-\,\,\widetilde{\as} \,\xi}\,n_{\rm in}\Lb \gamma"\Rb
\eeq
One can see that for $e^{\widetilde{\as} \,\xi} \,N\Lb Y, \xi\Rb\,\,
=\,\,\bar{N}\Lb Y, \xi\Rb$ we have:
\beq \label{ECN3}
\frac{\partial^2\,\bar{N}\Lb Y, \xi\Rb}{\partial\,Y\,\,\partial\,\xi}\,\,=\,\,\,\widetilde{\as}\,\bar{N}\Lb Y, \xi\Rb~~~
\mbox{or in the integral form:}~~ \,\,\frac{\partial\,N\Lb Y, \xi\Rb}{\partial\,Y}\,\,=\,\,\widetilde{\as}\int^\xi d \xi'\, {\cal K}\Lb \xi,\xi'\Rb N\Lb Y, \xi'\Rb
\eeq
with ${\cal K}\Lb \xi,\xi'\Rb \,\,=\,\,e^{ -\,\widetilde{\as}\Lb
 \xi \,-\,\xi'\Rb}$.

 We assume that the main contributions stem from the region
 $x_{02}\,\ll\,x_{01}$ and/or $x_{12}\,\ll\,x_{01}$
 Bearing this in mind we can use \eq{K2} for the kernel ,
 which can be re-written in the form:
 \bea \label{ECN31}
 \frac{\bas}{2 \pi}\int \, \displaystyle{K\Lb \vec{x}_{01};\vec{x}_{02},\vec{x}_{12}\Rb}\,d^2 x_{02} \,&\rightarrow&
\,\frac{\bas}{2}\, \int^{x^2_{01}}_{1/Q^2_s(Y,b)} \frac{ d x^2_{02}}{x_{02}^2}\,{\cal K}\Lb x_{10}; x_{02} \Rb\,\,+\,\,
\frac{\bas}{2}\, \int^{x^2_{01}}_{1/Q^2_s(Y, b)} \frac{ d |\vec{x}_{01}  -
 \vec{x}_{02}|^2}{|\vec{x}_{01}  - \vec{x}_{02}|^2}\,{\cal K}\Lb x_{10};|\vec{x}_{01}  - \vec{x}_{02}|\Rb\,\nn\\
 &=&\,\,\frac{\bas}{2}\, \int^{\xi}_{0} d \xi_{02} {\cal K}\Lb \xi,\xi_{02}\Rb
  \,\,+\,\,\frac{\bas}{2}\, \int^{\xi}_{0} d \xi_{12}{\cal K}\Lb
 \xi,\xi_{12}\Rb\eea
 where $\xi_{ik}\,\,=\,\,\ln \Lb x^2_{ik}\,Q^2_s(Y)\Rb$.
 
 Using \eq{ECN31} we can re-write the general BK equation
 (see \eq{BK}) in the form:
\beq \label{ECN4}
\frac{\partial N \Lb Y, \xi \Rb}
{ \partial Y\,}\,\,=\,\, \bas \int^\xi d \xi' \,K\Lb \xi,\xi'\Rb \,N \Lb Y, \xi' \Rb\Bigg( 1\,\,-\,\, N \Lb Y, \xi \Rb\Bigg)
\eeq
Using  \eq{ECN3} we can re-write \eq{ECN4} in the form:
\beq \label{ECN5}
\frac{\partial N \Lb Y, \xi \Rb}
{ \partial Y\,}\,\,=\,\, \bas \int^\xi \,d\xi' e^{-\widetilde{\as}\,\Lb \xi\,-\,\xi'\Rb} \,N \Lb Y, \xi' \Rb\Bigg( 1\,\,-\,\, N \Lb Y, \xi \Rb\Bigg)
\eeq
Introducing $N\Lb Y, \xi\Rb\,=\,1\,\,-\,\,\exp\Lb - 
\Omega\Lb Y,\xi\Rb\Rb$ we reduce \eq{ECN5} to the following expression:

\beq \label{ECN6}
\frac{\partial \Omega \Lb Y, \xi \Rb}
{ \partial Y\,}\,\,=\,\, \widetilde{\as} \int^\xi \,d\xi' e^{-\widetilde{\as}\,\Lb \xi\,-\,\xi'\Rb} \,\Lb 1\,\,-\,\,e^{ - \Omega \Lb Y, \xi' \Rb}\Rb
\eeq
Differentiating \eq{ECN6} with respect to $\xi$ we obtain:
\beq \label{ECN7}
\frac{\partial^2 \Omega \Lb Y, \xi \Rb}
{ \partial Y\,\partial \xi}\,\,=\,\, \widetilde{\as} \Bigg( 1\,\,-\,\,e^{ - \Omega \Lb Y, \xi\Rb}\Bigg)
\,\,-\,\, \widetilde{\as}^2 \int^\xi \,d\xi' e^{-\widetilde{\as}\,\Lb \xi\,-\,\xi'\Rb} \,\Lb 1\,\,-\,\,e^{ - \Omega \Lb Y, \xi' \Rb}\Rb
\eeq
Plugging the last term in \eq{ECN7} from \eq{ECN6} we have:
\beq \label{ECN8}
\frac{\partial^2 \Omega \Lb Y, \xi \Rb}
{ \partial Y\,\partial \xi}\,\,+\,\widetilde{\as}\,\frac{\partial \Omega \Lb Y, \xi \Rb}
{ \partial Y}\,\,=\,\, \widetilde{\as} \Bigg( 1\,\,-\,\,e^{ - \Omega \Lb Y, \xi\Rb}\Bigg)
\eeq

Looking for the solution which has the geometric scaling behaviour,
 we re-write \eq{ECN8} in new variable $z\,\,=\,\,\xi_s\,\,+\,\,\xi$
 and it takes the form
\beq \label{ECN9}
\frac{\partial^2 \Omega \Lb Y, \xi \Rb}
{ \partial z^2}\,\,+\,\,\widetilde{\as}\,\frac{\partial \Omega \Lb Y, \xi \Rb}
{ \partial z}\,\,=\,\, \sigma \Bigg( 1\,\,-\,\,e^{ - \Omega\Lb z\Rb}\Bigg)
\eeq

\begin{figure}
\centering 
   \includegraphics[width=9cm]{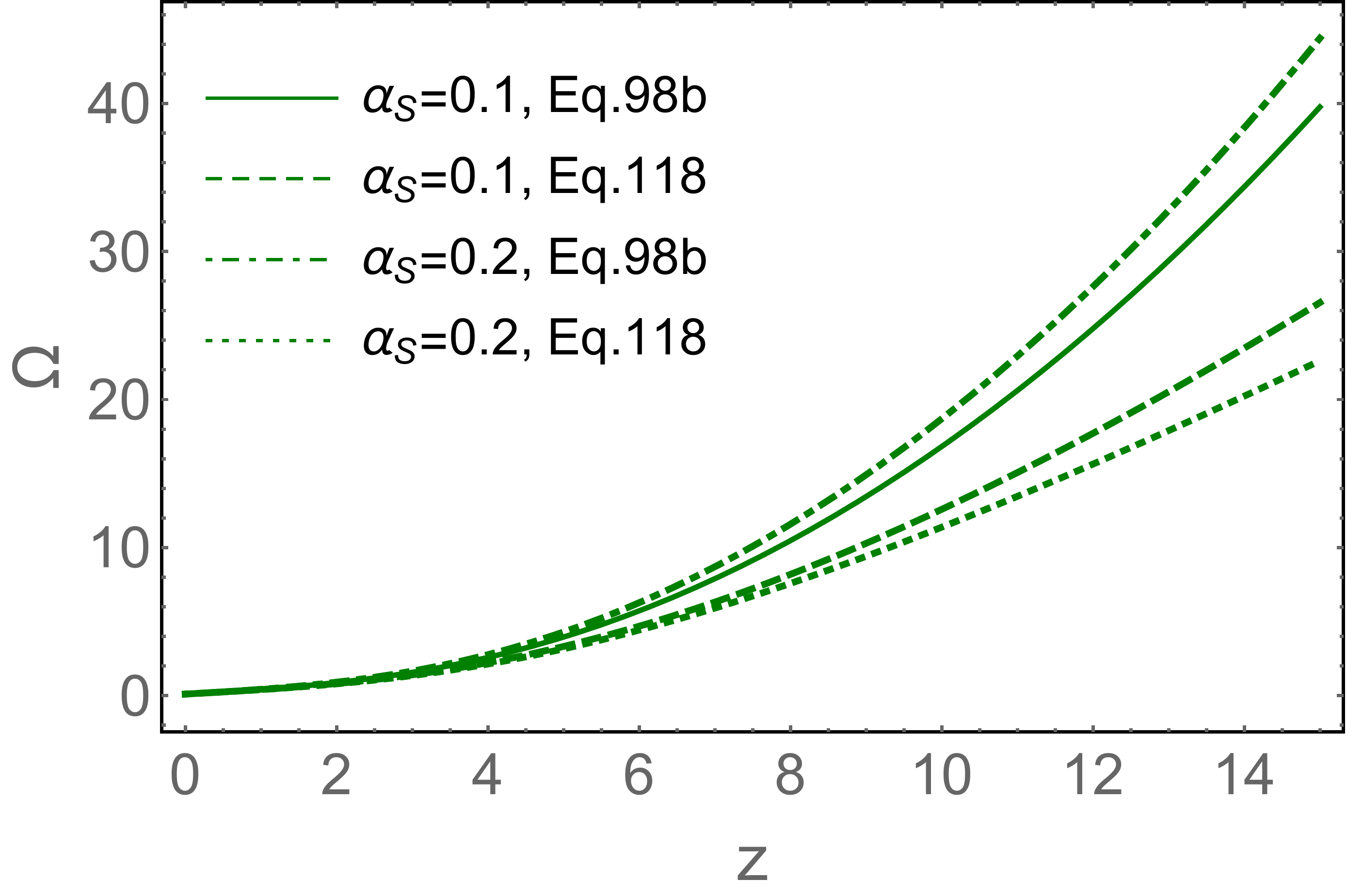}    
\caption{ Comparisons solutions to \eq{SOL2} and to \eq{ECN9} .}    
\label{comp}
\end{figure}

  One can see that deep in the saturation region, 
where $\Omega\,\,\gg\,\,1$,  $\Omega\Lb z\Rb\,\,=\,
\,\frac{1}{\lambda}\,z$. However, for $ z\,<\,1/\lambda$ we
 reproduce the solution of \eq{SOL8}. Therefore, we can infer
 that the energy conservation crucially changed the behaviour
 of the scattering amplitude  deep in the saturation region.
 In \fig{comp}
  we plot the solution to \eq{SOL2} and to \eq{ECN9}. One can
 see that the difference is large for  $z > 6$. At $z \,\leq\,5$,
 which corresponds the kinematic region of HERA, this difference
 is not so large and can be neglected.

  
   However, accounting of energy conservation is  beyond  the 
NLOBK,
 which is the subject of this paper. We believe, that it is necessary 
 to find
 the corrections of next-to-next-to leading order to treat the energy
 conservation on a theoretical footing.

\section{Conclusions}
  Concluding, we wish to formulate clearly the three stages of our approach:
  
  \begin{enumerate}
  \item\quad  For $ Q^2_s\,r^2\,\,\,<\,\,1$ (perturbative QCD region) 
  we suggest to use the linear evolution equation(see \eq{DLANL2}:
      \beq \label{MF1}
     \frac{\partial^2}{\partial \,\eta\,\partial\,\xi'} \,N\Lb \xi', \eta ; b \Rb\,\,+\,\, \frac{\partial}{\partial \,\eta}\,N\Lb \xi',  \eta ; b \Rb=\,\, \h\, \bas \,N\Lb \xi',  \eta ; b\Rb\,\,-\,\,\h\,\bas \frac{\partial}{\partial \,\xi'}\,N\Lb \xi',  \eta ; b \Rb    \eeq   
   We have discussed this equation in the section III-D-4. Recalling 
that in the  derivation of this equation we use the DLA in which
 both  $\eta\,\,=\,\,Y\,-\,\xi' $   and $\xi'$ are considered to
 be large: $ \bas\, \eta\, \xi \,\gg\,\,1$.  For $\eta \,>\,0$
 ($ Y\,>\,\xi$) we can use the experimental data as the initial
 condition for \eq{MF1}.    The value of the saturation momentum
 is given by \eq{ECDLA21}. Since its value turns out to be much
 less than $Q_{\rm max}$, which stems from the condition $\xi_{max}
 = \ln \Lb Q^2_{\rm max}/Q^2_0\Rb$\,\,=\,\,$\eta$, we can use the DGLAP evolution
 equation in the next-to-leading order for  $Q^2 \,>\,Q^2_{\rm max}$.

        \item\quad  For $ Q^2_s\,r^2\,\,\,\sim\,\,1$ (vicinity of the
 saturation scale) we use the scattering amplitude  in the form \cite{MUT}
        \beq \label{MF2}     
        N\Lb r,   \eta ; b \Rb \,\,=\,\,N_0\,\,\Lb  Q^2_s\Lb Y, b \Rb\,r^2\Rb^{\bar{\gamma}}
        \eeq
        with $\bar{\gamma}$ from \eq{ECDLA21}.
 \item\quad  For $ Q^2_s\,r^2\,\,\,\gg\,\,1$ (saturation region), we 
propose
 to use the solution to the non-linear equation which has the following form:
 \beq \label{MF3}
\int^{\Omega}_{\Omega_0}\frac{d \Omega'}{\sqrt{ \Omega^2_0\,\Lb1\,\,-\,\,\frac{\sigma}{\bar{\gamma}^2}\Rb\,\,+\,\,\frac{2\,\sigma}{\bar{\gamma}^2}\Lb -1\,\,+\,\,\Omega'\,\,+\,\,\,e^{ - \,\Omega'}\Rb
}}\,\,=\,\, \bar{\gamma}\,z \eeq 
  This equation does not take into account the corrections relating to 
 energy conservation, but it is simple, and for large region of $z$ the
 solution to the non-linear equation of section IV-D-1,  is described
 quite well by \eq{MF3}.
 \end{enumerate} 
 
     In general, we developed the approach based on the DLA approximation
 of perturbative QCD, and on the non-linear evolution for the leading
 twist approximation. We considered the non-linear equations, which
 arise in different kinematic regions , and discussed  solutions to them. 
  
  We believe that our suggested  approach which takes into account
 the main features of the NLO corrections to the BFKL kernel, and 
which  is likely to describe the experimental data for DIS
 processes. Indeed, \fig{lam}-b shows that the energy dependence
 of the saturation scale, as well as the value of $\bar{\gamma}$, are
 very close to the values that stem from  saturation models, that describe
 the data quite well (see Ref.\cite{RESH} for example).

\section{Acknowledgements}
   We thank our colleagues at Tel Aviv university and UTFSM for
 encouraging discussions. Our special thanks go  to E. Gotsman  for all 
his remarks and suggestions on this paper.  This research was supported  by 
 ANID PIA/APOYO AFB180002 (Chile),  Fondecyt (Chile) grants  
 1180118 and 1191434,  Conicyt Becas (Chile)  and PIIC 20/2020, DPP, Universidad T\'ecnica Federico Santa Mar\'ia.


\begin{thebibliography}{99}


 \bibitem{KOLEB}
Yuri V Kovchegov and Eugene Levin, {\it `` Quantum Choromodynamics at High Energies"}, Cambridge Monographs on Particle Physics, Nuclear Physics and Cosmology, Cambridge University Press, 2012 .
\bibitem{JIMWLK1}
J.~Jalilian-Marian, A.~Kovner, A.~Leonidov, and H.~Weigert, {\it ``The BFKL
  equation from the Wilson renormalization group"}
  \href{http://dx.doi.org/10.1016/S0550-3213(97)00440-9}, Nucl. Phys. {\bf
  B504} (1997)  415--431,
\href{http://arxiv.org/abs/hep-ph/9701284}[ arXiv:hep-ph/9701284].

\bibitem{JIMWLK2}
J.~Jalilian-Marian, A.~Kovner, A.~Leonidov, and H.~Weigert, {\it ``The Wilson
  renormalization group for low x physics: Towards the high density regime"}
  \href{http://dx.doi.org/10.1103/PhysRevD.59.014014}, Phys.Rev. {\bf D59}
  (1998)  014014,
\href{http://arxiv.org/abs/hep-ph/9706377}[arXiv:hep-ph/9706377
  [hep-ph]].

\bibitem{JIMWLK3}
A.~Kovner, J.~G. Milhano, and H.~Weigert, {\it ``Relating different approaches to
  nonlinear QCD evolution at finite gluon density"}
  \href{http://dx.doi.org/10.1103/PhysRevD.62.114005}, Phys. Rev. {\bf
  D62} (2000)  114005,
\href{http://arxiv.org/abs/hep-ph/0004014}[ arXiv:hep-ph/0004014].

\bibitem{JIMWLK4}
E.~Iancu, A.~Leonidov, and L.~D. McLerran, {\it Nonlinear gluon evolution in the
  color glass condensate. I"}
  \href{http://dx.doi.org/10.1016/S0375-9474(01)00642-X},Nucl. Phys. {\bf
  A692} (2001)  583--645,
\href{http://arxiv.org/abs/hep-ph/0011241}[ arXiv:hep-ph/0011241].

\bibitem{JIMWLK5}
E.~Iancu, A.~Leonidov, and L.~D. McLerran, {\it``The renormalization group
  equation for the color glass condensate"}
  \href{http://dx.doi.org/10.1016/S0370-2693(01)00524-X}, Phys. Lett. {\bf
  B510} (2001)  133--144,
\href{http://arxiv.org/abs/hep-ph/0102009}[ arXiv:hep-ph/0102009].
  \bibitem{BK}
I.~Balitsky,
{\it ``Operator expansion for high-energy scattering"},
[arXiv:hep-ph/9509348];\,\,
{\it ``Factorization and high-energy effective action"}, {\it Phys.\ Rev.} {\bf D60}, 014020 (1999)
[arXiv:hep-ph/9812311];\,\,\,\,
Y.~V.~Kovchegov,
{\it ``
Small-x $F_2$
structure function of a nucleus including multiple Pomeron exchanges"'}
{\it Phys.\ Rev.}  {\bf D60}, 034008  (1999),
[arXiv:hep-ph/9901281].
   
\bibitem{JIMWLK6}
E.~Ferreiro, E.~Iancu, A.~Leonidov, and L.~McLerran, {\it ``Nonlinear gluon
  evolution in the color glass condensate. II"}
  \href{http://dx.doi.org/10.1016/S0375-9474(01)01329-X}, Nucl. Phys. {\bf
  A703} (2002)  489--538,
\href{http://arxiv.org/abs/hep-ph/0109115}[ arXiv:hep-ph/0109115].

\bibitem{BFKL}
V.~S. Fadin, E.~A. Kuraev and L.~N. Lipatov,
{\it ``On the pomeranchuk singularity in asymptotically free theories"},
\newblock Phys. Lett. {\bf B60}, 50 (1975);\,\,\,
E.~A. Kuraev, L.~N. Lipatov and V.~S. Fadin,
{\it``The Pomeranchuk Singularity in Nonabelian Gauge Theories"}
\newblock Sov. Phys. JETP {\bf 45}, 199 (1977),
\newblock [Zh. Eksp. Teor. Fiz.72,377(1977)];\,\,\,
I.~I. Balitsky and L.~N. Lipatov,{\it ``The Pomeranchuk Singularity in Quantum Chromodynamics,''}
\newblock Sov. J. Nucl. Phys. {\bf 28}, 822 (1978),
\newblock [Yad. Fiz.28,1597(1978)].
\bibitem{LIP}
 L.~N.~Lipatov,
  {\it ``Small x physics in perturbative QCD,''}
  Phys.\ Rept.\  {\bf 286}, 131 (1997)
  [hep-ph/9610276];\,\,\,{\it ``The Bare Pomeron in Quantum Chromodynamics,''}
  Sov.\ Phys.\ JETP {\bf 63}, 904 (1986)
  [Zh.\ Eksp.\ Teor.\ Fiz.\  {\bf 90}, 1536 (1986)].
  \bibitem{NLOBK0}
I.~Balitsky,
  {\it ``Quark contribution to the small-x evolution of color dipole,''}
  Phys.\ Rev.\ D {\bf 75} (2007) 014001,
  [hep-ph/0609105].
  \bibitem{NLOBK01}
 Y.~V.~Kovchegov and H.~Weigert,
  {\it ``Triumvirate of Running Couplings in Small-x Evolution,''}
  Nucl.\ Phys.\ A {\bf 784} (2007) 188,
  [hep-ph/0609090].
  
  \bibitem{NLOBK1}
I.~Balitsky and G.~A. Chirilli, {\it ``Next-to-leading order evolution of color
  dipoles"}, \href{http://dx.doi.org/10.1103/PhysRevD.77.014019}
  Phys.Rev. {\bf D77} (2008)  014019,
\href{http://arxiv.org/abs/0710.4330}[arXiv:0710.4330 [hep-ph]].

\bibitem{NLOBK2}
I.~Balitsky and G.~A. Chirilli, {\it ``Rapidity evolution of Wilson lines at the
  next-to-leading order"},
  \href{http://dx.doi.org/10.1103/PhysRevD.88.111501} Phys.Rev. {\bf D88}
  (2013)  111501,
\href{http://arxiv.org/abs/1309.7644}[ arXiv:1309.7644 [hep-ph]].

\bibitem{JIMWLKNLO1}
A.~Kovner, M.~Lublinsky, and Y.~Mulian, {\it ``Jalilian-Marian, Iancu, McLerran,
  Weigert, Leonidov, Kovner evolution at next to leading order"},
  \href{http://dx.doi.org/10.1103/PhysRevD.89.061704} Phys.Rev. {\bf D89}
  (2014) no.~6, 061704,
\href{http://arxiv.org/abs/1310.0378}[ arXiv:1310.0378 [hep-ph]].

\bibitem{JIMWLKNLO2}
A.~Kovner, M.~Lublinsky, and Y.~Mulian, {\it ``NLO JIMWLK evolution unabridged"},
  \href{http://dx.doi.org/10.1007/JHEP08(2014)114} JHEP {\bf 08} (2014)
  114,
\href{http://arxiv.org/abs/1405.0418}[ arXiv:1405.0418 [hep-ph]].

\bibitem{JIMWLKNLO3}
M.~Lublinsky and Y.~Mulian, {\it ``High Energy QCD at NLO: from light-cone wave
  function to JIMWLK evolution"},
  \href{http://dx.doi.org/10.1007/JHEP05(2017)097} JHEP {\bf 05} (2017)
  097,
\href{http://arxiv.org/abs/1610.03453}[arXiv:1610.03453 [hep-ph]].
  \bibitem{DIMST}
B.~Duclou$\acute{e}$, E.~Iancu, A.~H.~Mueller, G.~Soyez and D.~N.~Triantafyllopoulos,
  {\it ``Non-linear evolution in QCD at high-energy beyond leading order,''}
  JHEP {\bf 1904} (2019) 081
  doi:10.1007/JHEP04(2019)081
  [arXiv:1902.06637 [hep-ph]] and references therein.
    \bibitem{BFKLNLO}
   V.~S.~Fadin and L.~N.~Lipatov,
  {\it ``BFKL pomeron in the next-to-leading approximation,''}
  Phys.\ Lett.\ B {\bf 429} (1998) 127
  [hep-ph/9802290].
  \bibitem{BFKLNLO1}\,M.~Ciafaloni and G.~Camici,
  {\it ``Energy scale(s) and next-to-leading BFKL equation,''}
  Phys.\ Lett.\ B {\bf 430} (1998) 349
  [hep-ph/9803389].
    \bibitem{SALAM}
G.~P.~Salam,
  {\it ``A Resummation of large subleading corrections at small x,''}
  JHEP {\bf 9807} (1998) 019
  doi:10.1088/1126-6708/1998/07/019
  [hep-ph/9806482];
  \bibitem{SALAM1}
   M.~Ciafaloni, D.~Colferai and G.~P.~Salam,
  {\it ``Renormalization group improved small $ x$  equation,''}
  Phys.\ Rev.\ D {\bf 60} (1999) 114036
  doi:10.1103/PhysRevD.60.114036
  [hep-ph/9905566].
  \bibitem{SALAM2}
 M.~Ciafaloni, D.~Colferai, G.~P.~Salam and A.~M.~Stasto,
  {\it ``Renormalization group improved small $x$ Green's function,''}
  Phys.\ Rev.\ D {\bf 68} (2003) 114003,
  [hep-ph/0307188].
      \bibitem{LETU}
E.~Levin and K.~Tuchin,
  {\it ``Solution to the evolution equation for high parton density QCD,''}
  Nucl.\ Phys.\ B {\bf 573}, 833 (2000)
  [hep-ph/9908317];\,\,\,
{\it ``New scaling at high-energy DIS,''}
  Nucl.\ Phys.\ A {\bf 691}, 779 (2001)
  [hep-ph/0012167]; {\it ``Nonlinear evolution and saturation for heavy nuclei in DIS,''}
   {\bf 693}, 787 (2001)
  [hep-ph/0101275].  
  
  
  \bibitem{GLR}
  L.~V.~Gribov, E.~M.~Levin and M.~G.~Ryskin,
  {\it ``Semihard Processes in QCD,''}
  Phys.\ Rept.\  {\bf 100} (1983) 1.
   \bibitem{MUQI}
A. H. Mueller and J. Qiu, 
{\it 
``Gluon recombination and shadowing at small values of $x$"},
Nucl. Phys. {\bf B268} (1986) 427
    \bibitem{MV}
L. McLerran and R. Venugopalan, 
{\it ``Computing quark and gluon distribution functions for very large nuclei"},
Phys. Rev. {\bf D49} (1994) 2233,
{\it ``Gluon distribution functions for very large nuclei at small transverse momentum"}, Phys. Rev. {\bf D49} (1994), 3352;     {\it `Green?s function in the color field of a large nucleus"}, {\bf D50} (1994) 2225; {\it ``
Fock space distributions, structure functions, higher twists, and small $x$"} ,
 {\bf D59} (1999) 09400.  
 
  
 
     \bibitem{KMRS}
  V.~A.~Khoze, A.~D.~Martin, M.~G.~Ryskin and W.~J.~Stirling,
  {\it ``The spread of the gluon $k_t$-distribution and the determination of the saturation scale at hadron colliders in resummed NLL BFKL,''}
  Phys.\ Rev.\ D {\bf 70} (2004) 074013
  [hep-ph/0406135].
  
 \bibitem{ASV} A.~Sabio Vera,
{\it ``An 'All-poles' approximation to collinear resummations in the Regge limit of perturbative QCD,''}
Nucl. Phys. B \textbf{722}, 65-80 (2005)
doi:10.1016/j.nuclphysb.2005.06.003
[arXiv:hep-ph/0505128 [hep-ph]].

           \bibitem{BALE}
J.~Bartels, E.~Levin,
  {\it ``Solutions to the Gribov-Levin-Ryskin equation in the nonperturbative region,''}
  Nucl.\ Phys.\  {\bf B387 } (1992)  617-637;  
   \bibitem{MUT}
A.~H.~Mueller and D.~N.~Triantafyllopoulos,
  {\it ``The Energy dependence of the saturation momentum,''}
  Nucl.\ Phys.\ B {\bf 640} (2002) 331
  [hep-ph/0205167]
  
      \bibitem{IIML}
   E.~Iancu, K.~Itakura and L.~McLerran,
  {\it ``Geometric scaling above the saturation scale,''}
  Nucl.\ Phys.\ A {\bf 708} (2002) 327
  [hep-ph/0203137].   
   
  \bibitem{SGBK}
 A.~M.~Stasto, K.~J.~Golec-Biernat, J.~Kwiecinski,
  {\it ``Geometric scaling for the total gamma* p cross-section in the low x region,''}
  Phys.\ Rev.\ Lett.\  {\bf 86 } (2001)  596-599,
  [hep-ph/0007192].  

      \bibitem{CLM}
  C.~Contreras, E.~Levin and R.~Meneses,
  {\it ``BFKL equation in the next-to-leading order: solution at large impact parameters,''}
  Eur.\ Phys.\ J.\ C {\bf 79} (2019) no.10,  842
  [arXiv:1906.09603 [hep-ph]].
  
        \bibitem{RY}
I. Gradstein and I. Ryzhik, {\it  Table of Integrals, Series, and Products},
Fifth Edition, Academic Press, London, 1994.  

\bibitem{DIFEQ}
Morris W. Hirsch, Stephen Smale and  Robert L. Devaney,{\it 
``Differential Equations, Dynamical Systems, and an Introduction to Chaos"},
 3rd Edition,Academic Press, 2013.
 
  
  \bibitem{MATH} A.D.  Polyanin and V.F. Zaitsev {\it ``Handbook  of nonlinear  partial differential equations"}, Chapman and Hall/CRC Press, 2004, Raca Baton, New York, London, Tokyo. 
  
   \bibitem{CLMP} 
  C.~Contreras, E.~Levin, R.~Meneses and I.~Potashnikova,
{\it ``CGC/saturation approach: a new impact-parameter dependent model in the next-to-leading order of perturbative QCD,''}
Phys. Rev. D \textbf{94} (2016) no.11, 114028
doi:10.1103/PhysRevD.94.114028
[arXiv:1607.00832 [hep-ph]].
  \bibitem{XCWZ}
W.~Xiang, Y.~Cai, M.~Wang and D.~Zhou,
{\it ``Rare fluctuations of the $S$-matrix at NLO in QCD,''}
Phys. Rev. D \textbf{99} (2019) no.9, 096026
doi:10.1103/PhysRevD.99.096026
[arXiv:1812.10739 [hep-ph]].    
  
     \bibitem{RESH}
 A.~H.~Rezaeian and I.~Schmidt,
  {\it``Impact-parameter dependent Color Glass Condensate dipole model and new combined HERA data,''}
  Phys.\ Rev.\ D {\bf 88} (2013) 074016
  [arXiv:1307.0825 [hep-ph]].
      \end{thebibliography}
\end{document}